\documentclass[sigconf]{acmart}

\copyrightyear{2026}
\acmYear{2026}
\acmConference[RecSys '26]{20th ACM Conference on Recommender Systems}{September 28--October 2, 2026}{Minneapolis, MN, USA}

\AtBeginDocument{%
  }

\renewcommand\footnotetextcopyrightpermission[1]{}
\usepackage{booktabs}
\usepackage{multirow}
\usepackage{graphicx}
\usepackage{amsmath}
\usepackage{amssymb}
\usepackage{enumitem}
\usepackage[table]{xcolor}

\usepackage{algorithm}
\usepackage{algpseudocode}

\begin{document}

\title{RecRec: Latent Interests Recursive Reasoning for Sequential Recommendation}

\author{Wenhao Deng}
\email{w.deng.1@research.gla.ac.uk}
\affiliation{%
  \institution{University of Glasgow}
  \city{Glasgow}
  \country{United Kingdom}
}

\author{Junchen Fu}
\authornote{Corresponding author.}
\email{j.fu.3@research.gla.ac.uk}
\affiliation{%
  \institution{University of Glasgow}
  \city{Glasgow}
  \country{United Kingdom}
}

\author{Hanwen Du}
\email{du.1128@osu.edu}
\affiliation{%
  \institution{Ohio State University}
  \city{Columbus}
  \state{OH}
  \country{USA}
}

\author{Alexandros Karatzoglou}
\email{alexandros.karatzoglou@gmail.com}
\affiliation{%
  \institution{Amazon}
  \city{Barcelona}
  \country{Spain}
}

\author{Ioannis Arapakis}
\email{ioannis.arapakis@telefonica.com}
\affiliation{%
  \institution{Telef\'onica Scientific Research}
  \city{Barcelona}
  \country{Spain}
}

\author{Hangjun Guo}
\email{hangjung@andrew.cmu.edu}
\affiliation{%
  \institution{Carnegie Mellon University}
  \city{Pittsburgh}
  \state{PA}
  \country{USA}
}

\author{Kaiwen Zheng}
\email{k.zheng.1@research.gla.ac.uk}
\affiliation{%
  \institution{University of Glasgow}
  \city{Glasgow}
  \country{United Kingdom}
}

\author{Yongxin Ni}
\email{niyongxin@u.nus.edu}
\affiliation{%
  \institution{National University of Singapore}
  \city{Singapore}
  \country{Singapore}
}

\author{Joemon M. Jose}
\email{Joemon.Jose@glasgow.ac.uk}
\affiliation{%
  \institution{University of Glasgow}
  \city{Glasgow}
  \country{United Kingdom}
}

\renewcommand{\shortauthors}{Deng et al.}

\begin{abstract}
Sequential recommender systems rely on a single forward pass to encode user interaction histories and predict the next item.
Increasing inference-time computation through latent reasoning, with the model proceeding step by step before the final prediction, has been recently explored in sequential recommendation with promising results.
However, \emph{how to structure the reasoning process for sequential recommendation remains an open question.}
Existing approaches couple reasoning and prediction in a single $d$-dimensional state, limiting reasoning depth and often relying on multi-stage pipelines with reinforcement learning (RL).
We propose \textbf{RecRec} (Recursive Reasoning for Recommendation), an RL-free framework that decouples reasoning from prediction, overcoming the fixed $d$-dimensional state bottleneck of prior methods.
RecRec consists of a Context Compressor and a Recursive Reasoner, trained in two simple supervised stages.
The Context Compressor distills the backbone's hidden states into a small set of latent interests, with an Interest Diversity Regularizer encouraging each interest to capture a distinct aspect of user behavior.
The Recursive Reasoner then refines these interests by reasoning in a separate intermediate latent space.
Deep supervision lets the reasoning depth be freely adjusted at inference without retraining.
On four real-world datasets, RecRec outperforms state-of-the-art reasoning-enhanced methods, and on three of four datasets, gains extend past the training-time depth.
Our findings point to a decoupled, multi-vector recipe that unleashes latent reasoning from the single-state bottleneck of prior methods, suggesting reasoning-state structure as a design axis to explore further in sequential recommendation.

\end{abstract}

\maketitle


\begin{figure*}[t]
  \centering
  \includegraphics[width=\linewidth]{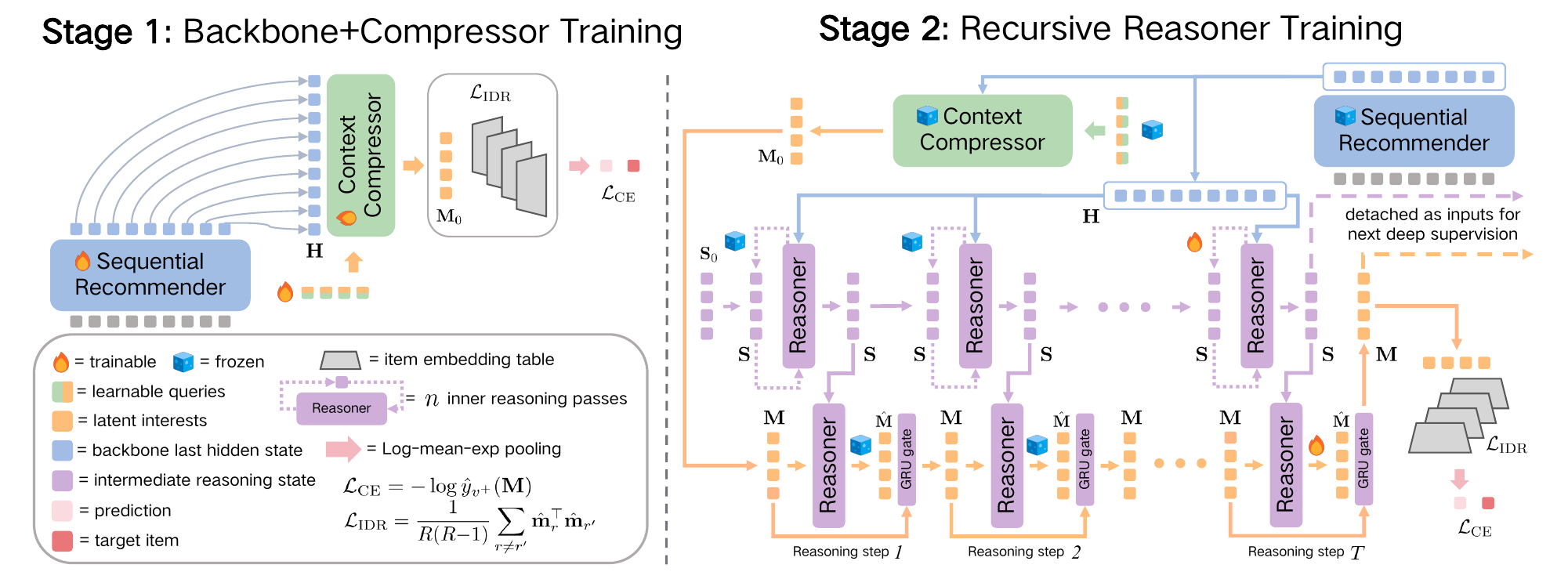}
  \caption{Training pipeline of RecRec.
    \textbf{Stage~1}: the backbone and Context Compressor are jointly trained, with $R$ learnable queries distilling the hidden states $\mathbf{H}$ into latent interests $\mathbf{M}_0$ that produce the next-item prediction.
    \textbf{Stage~2}: the backbone and Context Compressor are frozen; only the Recursive Reasoner is trained.
    The figure shows \emph{the first deep supervision round}, where $\mathbf{S}$ starts from $\mathbf{S}_0$ and $\mathbf{M}$ starts from $\mathbf{M}_0$.
    Within the round, the Reasoner executes $T$ reasoning steps.
    Each step updates $\mathbf{S}$ (middle, purple) via $n$ inner passes attending to $\mathbf{H}$ and the current $\mathbf{M}$, then commits an update to $\mathbf{M}$ (bottom, orange) through a GRU gate.
    Stage~2 repeats this round $K$ times; subsequent rounds inherit detached $\mathbf{S}$ and $\mathbf{M}$ from the previous round.}
  \label{fig:overview}
  \vspace{-0.8em}
\end{figure*}

\section{Introduction}

Sequential recommender systems encode a user's interaction history in a single forward pass and predict the next item from the resulting user representation.
While deeper or wider backbone encoders can improve this representation~\cite{zhai2024hstu}, the prediction is still produced in one shot, with no opportunity to revisit or refine the user's latent interests before committing to a ranking.
In large language models (LLMs), chain-of-thought reasoning~\cite{wei2022chain} and extended thinking enabled by reinforcement learning (RL)~\cite{shao2024deepseekmath,deng2025rapo,yu2025dapo,zheng2025gspo} have shown that allocating computation between the input and the final answer lets the model scratch and refine effectively through the intermediate reasoning steps, reducing errors that a single forward pass would commit to irreversibly.
Latent-space variants~\cite{hao2024coconut,geiping2025scaling} push this further by reasoning in continuous hidden states.
Bringing this idea to sequential recommendation has begun to show promising results~\cite{rearec2025,lares2025}.

How to structure the reasoning process for sequential recommendation, however, still remains an open question.
A central issue is that existing methods do not cleanly separate three distinct spaces: the user's behavior history (sequence), the intermediate reasoning, and the latent state used for prediction.
As summarized in Table~\ref{tab:spaces}, ReaRec~\cite{rearec2025} appends reasoning tokens directly to the backbone's input sequence, so reasoning shares the space with the history.
LARES~\cite{lares2025} iterates a shared block over all $L$ hidden states of the history itself, making reasoning a refinement of the sequence.
In both cases, the final prediction collapses to a single $d$-dimensional vector regardless of reasoning effort, a bottleneck on the prediction state that prior work on iterative architectures has identified as limiting modeling capacity~\cite{dao2024transformers,gu2022efficiently,bulatov2022recurrent}.

\begin{table}[!t]
  \centering
  \small
  \caption{Reasoning and prediction spaces across reasoning-enhanced sequential recommendation methods (the sequence space is the backbone's hidden states $\mathbf{H}$ in all cases).}
  \label{tab:spaces}
  \renewcommand{\arraystretch}{1.0}
  \begin{tabular*}{\columnwidth}{@{\extracolsep{\fill}}
    >{\raggedright\arraybackslash}p{1.3cm}
    >{\raggedright\arraybackslash}p{2.6cm}
    >{\raggedright\arraybackslash}p{2.3cm}
    c@{}}
    \toprule
    Method & Reasoning space & Pred.\ latent space & Decoupled? \\
    \midrule
    ReaRec & reasoning tokens appended to sequence & last reasoning token (or average) & No \\
    \cmidrule{1-4}
    LARES  & iterates over all $L$ hidden states   & last position of final iteration  & No \\
    \cmidrule{1-4}
    RecRec (ours) & dedicated reasoning state ($\mathbf{S}$) & dedicated latent interests ($\mathbf{M}$) & Yes \\
    \bottomrule
  \end{tabular*}
  \vspace{-1em}
\end{table}

A second issue is how reasoning depth is controlled.
ReaRec allocates reasoning position embeddings for a fixed number of steps; inference can use fewer steps than training, but cannot go beyond that number.
LARES allows the step count to vary at inference time, but achieving this flexibility requires randomizing the step count during training and an additional RL stage~\cite{shao2024deepseekmath}.

Recent recursive reasoning models~\cite{wang2025hierarchical,jolicoeur2025less} offer a different recipe for iterative latent computation.
The Tiny Recursion Model (TRM)~\cite{jolicoeur2025less} maintains two latent states updated by a single shared-weight block: a scratchpad for intermediate computation and an answer state that maps to the final output.
Deep supervision, applied at every detached step, teaches each step to improve whatever input it receives, so the model generalizes to more steps than it was trained on.
By recursively applying a single shared-weight block, TRM matches frontier LLMs on combinatorial tasks.
This simple yet effective dual-state, shared-weight, deep-supervision paradigm has not yet been explored for recommendation.

We propose \textbf{RecRec}, a backbone-agnostic framework that introduces the dual-state recursive reasoning paradigm to sequential recommendation.
A \emph{Context Compressor} distills the backbone's $L$ hidden states into a small set of $R$ latent interests.
A \emph{Recursive Reasoner} then alternates between refining an intermediate reasoning state, which revisits the user's behavior history and the current interests, and committing an update to the interests through a single Transformer block.
The framework is trained in two simple supervised stages.
Our main contributions are as follows.
\begin{itemize}[leftmargin=*,nosep]
\item We introduce RecRec, to the best of our knowledge the first framework to bring dual-state recursive reasoning to sequential recommendation.
Reasoning happens in a dedicated intermediate reasoning state that is separate from the latent interests used for prediction, so intermediate computation does not directly constrain the prediction state, removing the single $d$-dimensional bottleneck of prior methods.
\item A Recursive Reasoner bridges this paradigm to the recommendation setting with a single Transformer block alternating between the two states, where the intermediate reasoning state selectively attends to the user's behavior history and current interests.
This replaces TRM's concatenation-based MLP with attention-based information routing tailored to the recommendation setting.
\item A Context Compressor distills the backbone's hidden states into $R$ latent interests that jointly produce the prediction.
An \emph{Interest Diversity Regularizer} (IDR) encourages each interest to capture a distinct aspect of user behavior.
\item Through deep supervision with detached steps and without an RL stage, the reasoning depth can be freely adjusted at inference time beyond the reasoning depth used at training.
\end{itemize}

Experiments on four real-world datasets show that RecRec outperforms state-of-the-art reasoning-enhanced sequential recommenders.
On three of four datasets, gains extend past the training-time depth.

\section{Related Work}

\subsection{Sequential Recommendation}

Sequential recommendation predicts the next item a user will interact with based on their ordered interaction history.
GRU4Rec~\cite{hidasi2016session} and Caser~\cite{tang2018personalized} first applied recurrent and convolutional architectures to this task.
SASRec~\cite{kang2018self} introduced unidirectional self-attention that lets each position attend to all previous items.
BERT4Rec \cite{sun2019bert4rec} adopted bidirectional self-attention with a cloze-style masked item prediction objective, allowing each position to leverage both past and future context during training.
These Transformer-based models have become the de facto backbone for sequential recommendation, with later variants further incorporating modality features~\cite{yuan2023go,hou2022towards,li2023recformer,li2025exploring,fu2024iisan,fu2025efficient,fu2024exploring,ye2026multimodal,zhuang2025frequency,fu2025crossan,he2025double}.
Despite architectural differences, all of the above models share one property: they produce a single user representation from one forward pass through, and the prediction is read directly from this representation.
There is no mechanism to revisit or refine the user representation before prediction.

Several orthogonal lines of work relax this single-vector design: multi-interest recommenders replace the user vector with a set of interest vectors~\cite{li2019mind, cen2020comirec, zhang2022re4}, LLM-based methods cast recommendation as language modeling or align LLMs with backbones~\cite{geng2022p5, bao2023tallrec, liao2024llara, li2024e4srec}, and generative retrieval replaces dense scoring with semantic-ID autoregression~\cite{rajput2023tiger, fu2026diger}.

\subsection{Reasoning-Enhanced Sequential Recommendation}

Recent work increases inference-time computation for sequential recommendation by introducing multi-step latent reasoning on top of a backbone encoder.

ReaRec~\cite{rearec2025} was the first to bring latent reasoning into sequential recommendation.
It reuses the backbone for several forward passes, feeding the last hidden state back through the same encoder with dedicated reasoning position embeddings.
Each step appends reasoning tokens to the input, so the attention cost per step grows as reasoning tokens accumulate.
The reasoning position embeddings are allocated for a fixed number of steps, so inference can use fewer steps than training but cannot use more steps than at training.
ReaRec trains with either Ensemble Reasoning Learning (ERL), which aggregates reasoning states, or Progressive Reasoning Learning (PRL), which applies per-step supervision with progressively sharpened targets.
LARES~\cite{lares2025} also reasons over the full sequence, applying a shared core-block $k$ times, with $k$ sampled during training, to allow flexible depth at test time.
LARES trains in three stages: self-supervised pre-training with trajectory-level and step-level alignment, followed by reinforcement post-training via GRPO using ranking metrics as reward signals.

From an architectural perspective, these two methods represent opposite ends of the evolving state design space.
ReaRec evolves a single $d$-dimensional vector per step.
LARES evolves all $L$ hidden states at every step through full self-attention, yet only the last position of the final iteration is used for prediction.
Both therefore collapse to a single $d$-dimensional vector at the prediction step.
In both cases the state being refined is also the state from which the prediction is made, with no separate space for intermediate reasoning.
RecRec takes a middle ground by compressing the backbone's output into $R$ latent interests and reasoning exclusively on them.

Several concurrent efforts scale reasoning along complementary axes.
Parallel Latent Reasoning~\cite{tang2026parallel} scales reasoning width via aggregated parallel streams within a single-state paradigm, complementary to our depth-extrapolation axis.
Diffusion-based two-stage refinement~\cite{jiang2026diffureason}, manifold-constrained adaptive test-time compute~\cite{yang2026mancar}, and process-reward-guided beam search in generative retrieval~\cite{guo2026promise} explore further orthogonal directions.
Prompt-level reflection frameworks over LLM backbones~\cite{qin2025more} operate in a different paradigm from latent-space reasoning.
Across these directions, RecRec's reason-then-update loop, inspired by TRM~\cite{jolicoeur2025less} (discussed next), is distinguished by maintaining a dedicated intermediate reasoning state separate from the evolving interests.

\subsection{Recursive Reasoning Models}
\label{sec:rw_recursive}

Chain-of-thought prompting~\cite{wei2022chain} demonstrated that multi-step reasoning substantially improves LLM performance on tasks requiring compositional logic, and recent reasoning-focused models~\cite{guo2025deepseek} have scaled this idea with RL.
A parallel line of work replaces discrete token generation with continuous latent reasoning: Coconut~\cite{hao2024coconut} trains LLMs to reason entirely in continuous hidden states without emitting intermediate tokens, and recurrent-depth models~\cite{geiping2025scaling} apply depth-wise recursion in latent space to scale test-time computation.
The Universal Transformer~\cite{dehghani2019universal} earlier showed that iteratively applying the same Transformer block improves compositional generalization.
Generating explicit reasoning tokens is expensive and bottlenecked by the vocabulary; latent reasoning can iterate freely in a continuous representation space.

The Hierarchical Reasoning Model (HRM)~\cite{wang2025hierarchical} introduced a two-module recurrent architecture where a high-level module maintains a slow planning state and a low-level module handles rapid step-by-step computation.
Building on HRM, TRM~\cite{jolicoeur2025less} unifies both modules into a single shared-weight block that alternates between two roles.
An answer state~$y$ maps to the predicted output, while a scratchpad state~$z$ holds intermediate reasoning but is never decoded.
Within each round the block is applied $n$~times to update~$z$, then once to update~$y$.
Training uses $T{-}1$ no-gradient warmup rounds followed by one back-propagated round (1-step BPTT), repeated across $K$~detached deep supervision steps so each step independently learns to improve whatever input it receives.
Because a single block handles all computation, TRM achieves strong results on combinatorially hard tasks such as Sudoku and ARC-AGI.

RecRec introduces recursive reasoning to sequential recommendation.
The latent interests~$\mathbf{M}$ correspond to TRM's answer state~$y$, and the intermediate reasoning state~$\mathbf{S}$ to the scratchpad~$z$.

\section{Method}
\label{sec:method}

RecRec is designed to work with any sequential backbone.
Figure~\ref{fig:overview} illustrates the overall architecture.
Given a sequential backbone, a Context Compressor~(\S\ref{sec:ic}) first extracts the last layer's hidden states into a compact array of $R$ latent interests via learned queries and cross-attention.
A Recursive Reasoner~(\S\ref{sec:reasoner}) then iteratively refines these interests through a reason-then-update loop.
The refined interests directly predict item relevance via log-mean-exp pooling~(\S\ref{sec:prediction}).
The framework is trained in two supervised stages~(\S\ref{sec:training}).

\subsection{Problem Formulation}
\label{sec:problem}

In sequential recommendation, each user is associated with a chronologically ordered interaction sequence $\mathbf{x} = (v_1, v_2, \ldots, v_L)$ where $v_i \in \mathcal{V}$ denotes an item from the item set.
The task is to predict the next item $v_{L+1}$ the user is likely to interact with.

A sequential backbone (e.g., SASRec~\cite{kang2018self} or BERT4Rec~\cite{sun2019bert4rec}) encodes the interaction sequence and produces hidden states $\mathbf{H} \in \mathbb{R}^{L \times d}$, where $L$ is the sequence length and $d$ is the hidden dimension.
Standard practice predicts from only the final position's hidden state~$\mathbf{h}_L$, compressing the entire interaction history into a single $d$-dimensional vector.

RecRec augments this pipeline by extracting $R$ latent interests from~$\mathbf{H}$, iteratively refining them through recursive reasoning, and predicting items directly from the refined interests.
Given the user's behavior history~$\mathbf{H}$ and an initial estimate of their interests, the question RecRec asks at each reasoning step (formally defined in \S\ref{sec:reasoner}) is: with what is  known now, how can these interests be revised to better anticipate what the user wants next?

\subsection{Context Compressor}
\label{sec:ic}

The Context Compressor extracts the backbone's last-layer hidden states~$\mathbf{H} \in \mathbb{R}^{L \times d}$ into a compact array of $R$ latent interests $\mathbf{M}_0 \in \mathbb{R}^{R \times d}$, reducing a variable-length interaction history to a fixed-size representation for the downstream Recursive Reasoner to refine.
The compressor uses $R$ learned queries that cross-attend to the input sequence and are refined through cross-attention layers, following the learned-query cross-attention design~\cite{jaegle2021perceiver,carion2020detr}.

The input sequence is first normalized once: $\bar{\mathbf{H}} = \mathrm{LN}_{\text{input}}(\mathbf{H})$, and this normalized representation is shared across all layers.
The learned queries are initialized with Xavier uniform  and expanded to the batch dimension; no positional encoding is applied, as the $R$ latent interests form an unordered set.

Let $\mathbf{Z}$ denote the queries being refined, initialized to the learned queries.
Each layer applies three pre-norm residual sub-operations:
\begin{align}
  \mathbf{Z} &\leftarrow \mathbf{Z}
    + \mathrm{CrossAttn}\!\bigl(
        \mathrm{LN}_{\text{crossattn}}(\mathbf{Z}),\;
        \bar{\mathbf{H}},\;
        \bar{\mathbf{H}}
      \bigr),
  \label{eq:ic_cross} \\[3pt]
  \mathbf{Z} &\leftarrow \mathbf{Z}
    + \mathrm{SelfAttn}\!\bigl(
        \mathrm{LN}_{\text{selfattn}}(\mathbf{Z})
      \bigr),
  \label{eq:ic_self} \\[3pt]
  \mathbf{M}_0 &= \mathbf{Z}
    + \mathrm{FFN}\!\bigl(
        \mathrm{LN}_{\text{ffn}}(\mathbf{Z})
      \bigr).
  \label{eq:ic_ffn}
\end{align}
In the cross-attention, normalization is applied to the queries; $\bar{\mathbf{H}}$ reuses the input normalization.
Padding positions in $\bar{\mathbf{H}}$ are masked via a key padding mask.
Neither cross-attention nor self-attention uses a causal mask, as the $R$ latent interests form an unordered set.

\subsection{Recursive Reasoner}
\label{sec:reasoner}

\begin{figure}[!t]
  \centering
  \includegraphics[width=\linewidth]{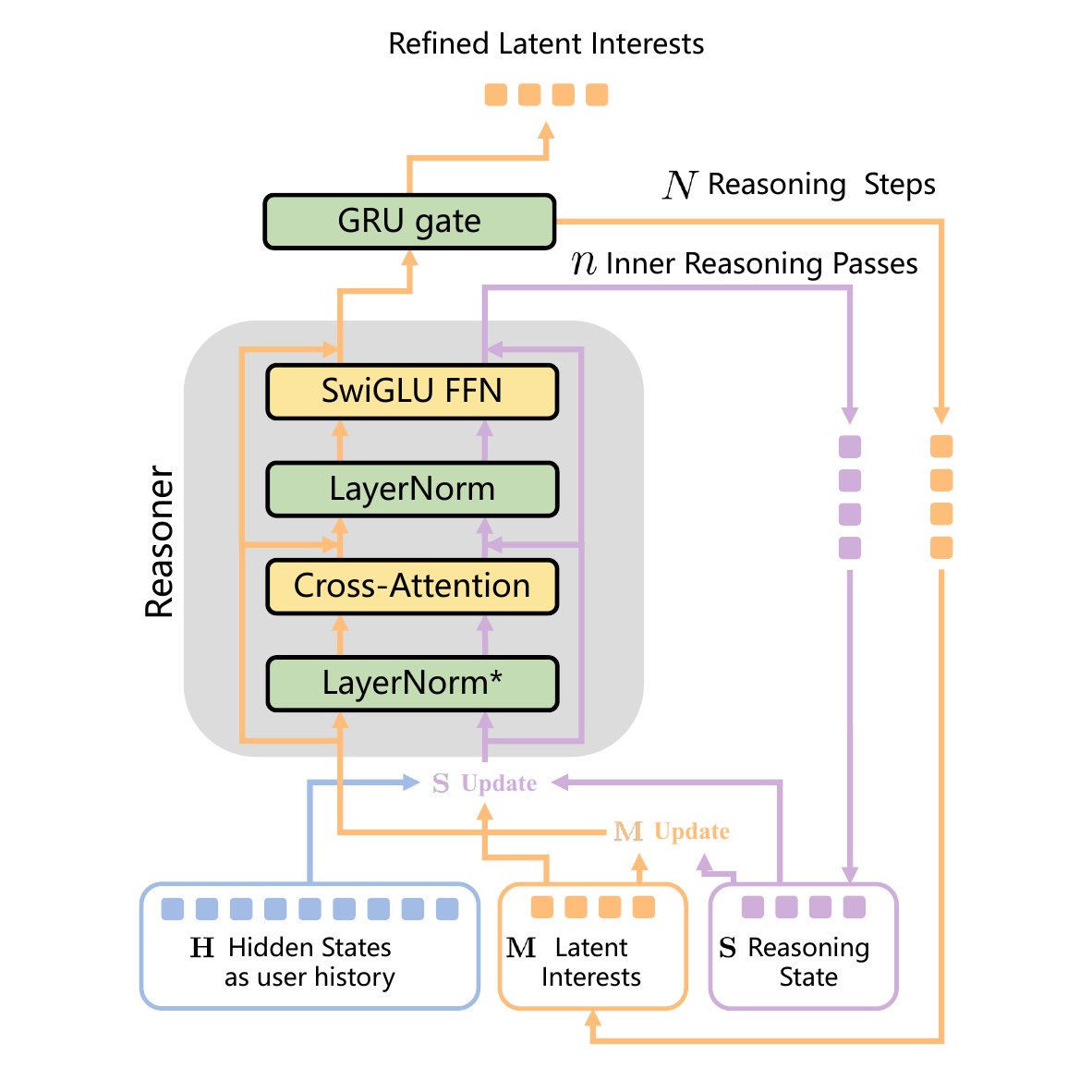}
  \caption{Recursive Reasoner architecture and inference behavior.
    The block is applied to both states in alternation, not in parallel.
    In the $\mathbf{S}$ update phase, $\mathbf{S}$ attends to $\mathbf{H}$ and the current $\mathbf{M}$; in the $\mathbf{M}$ update phase, $\mathbf{M}$ attends to the refined $\mathbf{S}$ and is committed through a GRU gate to produce the refined latent interests.
    Each reasoning step first performs $n$ inner $\mathbf{S}$ updates, then one $\mathbf{M}$ update, and the outer loop runs $N$ reasoning steps ($N = K \times T$ at training; freely adjustable at inference).
    The asterisk on the first LayerNorm denotes the separate $\mathbf{Q}$ / external-$\mathbf{KV}$ normalization detailed in \S\ref{sec:reasoner}.}
  \label{fig:reasoner}
  \vspace{-1em}
\end{figure}

The Recursive Reasoner iteratively refines the latent interests $\mathbf{M}_0$ produced by the Context Compressor.
To decouple the roles of reasoning and prediction, we maintain two separate states.
\emph{Latent interests}~$\mathbf{M} \in \mathbb{R}^{R \times d}$ are the representations decoded into predictions; they are initialized from $\mathbf{M}_0$.
An \emph{intermediate reasoning state}~$\mathbf{S} \in \mathbb{R}^{R \times d}$ carries out intermediate reasoning but is never directly used for prediction.
We match $\mathbf{S}$ to $\mathbf{M}$'s shape so that a single shared Reasoner block (introduced below) processes both with the same query dimension.
A \emph{reasoning step} consists of $n$ inner passes in which $\mathbf{S}$ gathers information from $\mathbf{H}$ and the current $\mathbf{M}$, followed by one gated update that commits this information back to $\mathbf{M}$ through a GRU~\cite{cho2014learning} gate.
At the start of each supervision round, $\mathbf{S}$ is initialized from a fixed random tensor $\mathbf{S}_0 \in \mathbb{R}^{R \times d}$, drawn once at model construction from a truncated normal distribution.
We keep $\mathbf{S}_0$ fixed rather than learnable because the first $T{-}1$ reasoning steps of every supervision round run without gradient (see below), so a learnable $\mathbf{S}_0$ would never receive a gradient signal to update.
This separation of roles lets the model reason deeply through $\mathbf{S}$ without destabilizing $\mathbf{M}$.
In particular, since $\mathbf{S}$ is never used for prediction, its representations can evolve at a different magnitude from the latent interests, which must stay compatible with the item embeddings used to predict the relevance (via dot product, see~\S\ref{sec:prediction}).
The GRU gate on $\mathbf{M}$ further controls how much of this freely evolved reasoning is committed to the prediction state at each step.
A single Transformer block handles both roles, updating $\mathbf{S}$ and $\mathbf{M}$ with the same parameters but different inputs.
The two-state formulation follows recursive reasoning models~\cite{jolicoeur2025less}.

\paragraph{The Reasoner block.}
Both states are updated by a single Transformer~\cite{vaswani2017attention} block, denoted as $\mathrm{Reasoner}(\cdot,\cdot)$ in the equations below, with pre-norm LayerNorm, bias-free multi-head attention, and a SwiGLU~\cite{shazeer2020glu} feed-forward network.
Queries and key-value inputs are normalized by separate LayerNorm layers, matching the normalization structure of the Context Compressor.
When additional KV tokens~$\mathbf{C}$ are provided, they are normalized by a separate KV norm and appended to the key-value set:
\begin{align}
  \mathbf{X}' &= \mathbf{X}
    + \mathrm{Attn}\!\bigl(
        \mathrm{LN}_q(\mathbf{X}),\;
        [\mathrm{LN}_q(\mathbf{X});\, \mathrm{LN}_{kv}(\mathbf{C})]
      \bigr),
  \label{eq:swb_attn} \\
  \mathrm{Reasoner}(\mathbf{X}, \mathbf{C})
    &= \mathbf{X}'
    + \mathrm{FFN}\!\bigl(\mathrm{LN}_{\text{ffn}}(\mathbf{X}')\bigr),
  \label{eq:swb_ffn}
\end{align}
where $\mathrm{LN}_q$, $\mathrm{LN}_{kv}$, and $\mathrm{LN}_{\text{ffn}}$ are separate LayerNorm layers.
Both the attention output projection and the FFN output layer are zero-initialized, so the Reasoner starts as an identity mapping.
This stabilizes early training by letting the Reasoner learn to modify the representations it receives from the Context Compressor.

\paragraph{The reasoning step.}
Each reasoning step proceeds in two phases.
First, the intermediate reasoning state is updated for $n$ inner reasoning passes, each time attending to the full backbone hidden states $\mathbf{H}$ and the current latent interests:
\begin{equation}
  \mathbf{S} \leftarrow
    \mathrm{Reasoner}\bigl(\mathbf{S},\; [\mathbf{H};\, \mathbf{M}]\bigr),
  \quad \text{repeated } n \text{ times},
  \label{eq:s_update}
\end{equation}
where $\mathbf{H} \in \mathbb{R}^{L \times d}$ is the sequence of hidden states from the frozen backbone.
Padding positions in $\mathbf{H}$ are masked in the key-value attention; $\mathbf{M}$ is held fixed while $\mathbf{S}$ evolves, so each pass lets $\mathbf{S}$ re-examine the encoded user history in light of the current interests.

Second, $\mathbf{M}$ reads the refined intermediate reasoning state through the Reasoner block, producing a \emph{candidate update}~$\hat{\mathbf{M}}$ that a GRU gate then blends into the committed $\mathbf{M}$:
\begin{align}
  \hat{\mathbf{M}} &=
    \mathrm{Reasoner}(\mathbf{M},\; \mathbf{S}),
    \label{eq:m_read} \\
  \mathbf{M} &\leftarrow
    \mathrm{GRU}(\hat{\mathbf{M}},\; \mathbf{M}_{\mathrm{prev}}),
    \label{eq:m_gate}
\end{align}
where $\mathbf{M}_{\mathrm{prev}}$ is the value of $\mathbf{M}$ before the current reasoning step.
Directly overwriting $\mathbf{M}$ with $\hat{\mathbf{M}}$ risks large jumps per reasoning step.
Following Slot Attention~\cite{locatello2020object}, we apply a GRU~\cite{cho2014learning} gate that blends $\hat{\mathbf{M}}$ with $\mathbf{M}_{\mathrm{prev}}$, letting the model learn how much to change at each step.
The gate's inputs are layer-normalized before the blending ratio is computed, so the update decision is not affected by representation magnitude.

\paragraph{Deep supervision and reasoning depth.}
Rather than training the Reasoner with a single loss at the end of all reasoning, we adopt deep supervision~\cite{wang2025hierarchical}: a loss is computed after each of $K$ supervision rounds, with $\mathbf{S}$ and $\mathbf{M}$ detached between rounds.
This detachment forces each round to improve whatever state it receives, rather than co-adapting with earlier rounds through long gradient chains: each round learns the same ``given state, make it better'' objective independently.
Each supervision round executes $T$ reasoning steps, of which only the final step is back-propagated through; the first $T{-}1$ steps execute without gradient, preparing a good starting point for $\mathbf{M}$ without consuming gradient memory (a 1-step back-propagation-through-time (BPTT) strategy, following~\cite{jolicoeur2025less}).
The total training reasoning depth is therefore $K \times T$ steps, each containing $n$ inner reasoning passes.
Because each round operates on its inputs independently with no dependence on the history of preceding rounds, the Reasoner can be applied for additional steps beyond the training configuration at inference, deepening the refinement of the latent interests before prediction.

\subsection{Prediction}
\label{sec:prediction}

After reasoning, the refined latent interests $\mathbf{M}_K \in \mathbb{R}^{R \times d}$ are used to predict the relevance of all candidate items.
Each latent interest computes a per-item relevance via inner product with the item embedding $\mathbf{e}_v \in \mathbb{R}^d$ from the shared embedding table, scaled by a temperature $\tau$:
\begin{equation}
  s_r(v) = \frac{\mathbf{m}_r^\top \mathbf{e}_v}{\tau},
  \quad r = 1, \ldots, R.
  \label{eq:per_interest_score}
\end{equation}
The $R$ per-interest relevances are aggregated into a single prediction $s(v)$.
We want the most relevant interest to dominate the prediction, while keeping gradients flowing through all $R$ interests so that every interest can be refined during training.
A log-mean-exp pooling provides this soft maximum behavior:
\begin{equation}
  s(v)
  = \log \frac{1}{R} \sum_{r=1}^{R} \exp\bigl(s_r(v)\bigr).
  \label{eq:lse_pool}
\end{equation}
This serves as the final prediction for both training and evaluation.

\subsection{Training and Inference}
\label{sec:training}

RecRec is trained in two supervised stages.

\paragraph{Stage~1: Backbone and compressor co-training.}
The backbone and Context Compressor are jointly trained from scratch.
The backbone encodes the interaction sequence into $\mathbf{H}$, and the Context Compressor produces initial latent interests~$\mathbf{M}_0$.
The Recursive Reasoner is not used in this stage; item relevance is predicted directly from $\mathbf{M}_0$ via log-mean-exp pooling (Eq.~\eqref{eq:lse_pool}) and supervised with cross-entropy loss against the ground-truth next item $v^+$.
An interest diversity regularizer $\mathcal{L}_{\text{IDR}}$ is added to prevent the learned queries from collapsing to identical interests.
The softmax probability of the ground-truth item under the scores $s(\cdot)$ (Eq.~\eqref{eq:lse_pool}) computed from a given interest tensor is:
\begin{equation}
  \hat{y}_{v^+}(\mathbf{M})
  = \frac{\exp\bigl(s(v^+)\bigr)}
         {\sum_{v \in \mathcal{V}} \exp\bigl(s(v)\bigr)},
  \label{eq:softmax_prob}
\end{equation}
where the scores $s(v)$ are obtained from $\mathbf{M}$ via Eq.~\eqref{eq:per_interest_score} and Eq.~\eqref{eq:lse_pool}.
The Stage~1 loss is:
\begin{equation}
  \mathcal{L}_1
  = -\log \hat{y}_{v^+}(\mathbf{M}_0)
  + \lambda_{\text{IDR}} \cdot \mathcal{L}_{\text{IDR}}.
  \label{eq:loss_s1}
\end{equation}

\paragraph{Interest diversity regularization (IDR)}
All $R$ learned queries attend to the same hidden states~$\mathbf{H}$ via cross-attention.
Without explicit regularization, they converge to attend to the same region of the sequence, producing near-identical latent interests (inter-interest cosine similarity exceeds $0.8$ in practice).
IDR penalizes this collapse by minimizing the mean off-diagonal cosine similarity among $\mathbf{M}_0$'s interests:
\begin{equation}
  \mathcal{L}_{\text{IDR}}
  = \frac{1}{R(R{-}1)}
    \sum_{r \neq r'}
    \hat{\mathbf{m}}_r^\top \hat{\mathbf{m}}_{r'},
  \label{eq:idr_loss}
\end{equation}
where $\hat{\mathbf{m}}_r = \mathbf{m}_r / \lVert \mathbf{m}_r \rVert$ denotes the L2-normalized interest.
With IDR active, inter-interest cosine drops from $0.82$--$0.90$ to $0.14$--$0.75$ across datasets (see Figure~\ref{fig:idr_mechanism}), maintaining sufficient diversity for the downstream prediction to benefit from distinct interests.

\paragraph{Stage~2: Recursive Reasoner training.}
The backbone is frozen (eval mode).
The Context Compressor is also frozen but keeps dropout active, serving as a regularizer on the Reasoner's input.
The Recursive Reasoner is initialized from scratch and is the only trainable component, zero-initialized so that it starts as an identity mapping, stabilizing early training.

Training follows the deep supervision scheme of \S\ref{sec:reasoner} with $K$ supervision rounds per batch.
At each round $k$: the Reasoner takes $\mathbf{H}$ and the detached $\mathbf{S}$, $\mathbf{M}$ from the previous round and runs $T$ reasoning steps; only the final step, consisting of $n$ inner reasoning passes on $\mathbf{S}$ and one gated update on $\mathbf{M}$, is back-propagated through (1-step BPTT); the loss in Eq.~\eqref{eq:loss_s2} is computed on the resulting $\mathbf{M}^{(k)}$, one gradient step is taken, and both $\mathbf{S}^{(k)}$ and $\mathbf{M}^{(k)}$ are detached before being fed to round $k{+}1$.
At the first round, $\mathbf{S}$ is initialized from $\mathbf{S}_0$ and $\mathbf{M}$ from $\mathbf{M}_0$.
The Stage~2 loss at round $k$ adds IDR to the cross-entropy on $\mathbf{M}^{(k)}$:
\begin{equation}
  \mathcal{L}_2^{(k)} =
  -\log \hat{y}_{v^+}(\mathbf{M}^{(k)})
  + \lambda_{\text{IDR}} \cdot \mathcal{L}_{\text{IDR}}(\mathbf{M}^{(k)}),
  \quad k = 1, \ldots, K.
  \label{eq:loss_s2}
\end{equation}

\paragraph{Inference.}
Figure~\ref{fig:reasoner} summarizes the Reasoner block and its outer loop at inference.
At inference, the Reasoner is applied for $N$ reasoning steps, each consisting of $n$ inner passes on $\mathbf{S}$ and one gated update to $\mathbf{M}$.
At training, $N = K \times T$ is determined by the deep supervision schedule (Eq.~\eqref{eq:loss_s2}); at inference, there is no supervision, so $N$ can be chosen freely.
Because the Reasoner learns to improve whatever state it receives, the model can be evaluated at values of $N$ larger than used during training, without retraining.

\section{Experiments}

\begin{table}[!t]
\caption{Statistics of the experimental datasets.}
\label{tab:datasets}
\begin{tabular}{lrrrrc}
\toprule
Dataset & \#Users & \#Items & \#Inter. & Avg.L & Sparsity \\
\midrule
S-Shop    &  89,284 & 15,552 &   384,741 & 12.64 & 99.97\% \\
V-Shop &  54,005 & 21,439 &   249,505 & 14.77 & 99.98\% \\
MicroLens   &  98,133 & 16,959 &   502,561 &  5.12 & 99.97\% \\
Steam       & 157,998 &  7,960 & 1,292,442 & 25.97 & 99.90\% \\
\bottomrule
\end{tabular}
\end{table}

We organize the experimental evaluation around three research questions: \textbf{RQ1:} Does RecRec's dual-state design outperform prior reasoning-enhanced methods? \textbf{RQ2:} Does deep supervision enable flexible reasoning depth at inference, including beyond the training-time depth? \textbf{RQ3:} How do the latent interests and other components of RecRec contribute to its overall performance?

\subsection{Experimental Setup}

\subsubsection{Datasets.}
We conduct experiments on four datasets: two shopping datasets~\cite{hou2024bridging}, a short-video dataset (MicroLens~\cite{ni2023content}), and a gaming dataset (Steam~\cite{kang2018self}), all filtered with a 5-core threshold.
For simplicity, we refer to the two shopping datasets as S-Shop and V-Shop.
To ensure training quality, we further remove users whose training split contains fewer than 5 interactions.
Statistics are in Table~\ref{tab:datasets}.

\textbf{S-Shop and V-Shop.} The two shopping datasets contain product ratings and reviews: S-Shop covers user interactions with software products, while V-Shop covers user interactions with video games.
Together they span e-commerce scenarios of different scales and sparsity levels.
To focus on positive interaction signals, we retain only records with ratings above 3.
Interaction sequences are constructed by sorting records by timestamp.
Since both datasets provide absolute timestamps, the training, validation, and test splits follow the official temporal partitions.

\textbf{MicroLens-100K.} MicroLens-100K is a short-video recommendation dataset that records user click behavior on a short-video platform.
Unlike the e-commerce datasets, it contains only implicit feedback (clicks), making the interaction patterns closer to real-world feed recommendation scenarios.
The raw data is provided as time-ordered user interaction sequences without absolute timestamps.
We adopt a relative time split at the 80\% and 90\% positions of each user's sequence length for training, validation, and test sets.

\textbf{Steam.} The Steam dataset is collected from user reviews on the Steam gaming platform and serves as a classic benchmark in sequential recommendation, widely adopted by representative methods such as SASRec and BERT4Rec.
Since the raw data may contain multiple reviews from the same user for the same game, we deduplicate and keep only the first review.
As with MicroLens, the dataset lacks absolute timestamps, so we adopt a relative time split at the 80\% and 90\% positions of each user's interaction sequence.

\subsubsection{Baselines.}
We compare RecRec against two groups of baselines.
\noindent\textbf{Non-reasoning methods.}
\textbf{SASRec}~\cite{kang2018self} uses causal self-attention to encode user interaction sequences.
\textbf{BERT4Rec}~\cite{sun2019bert4rec} applies bidirectional self-attention with masked item prediction.

\noindent\textbf{Reasoning-enhanced methods.}
\textbf{ERL} and \textbf{PRL}~\cite{rearec2025} are two variants of ReaRec that perform multi-step reasoning by autoregressively feeding the last hidden state back into the backbone encoder.
ERL aggregates reasoning states via ensemble, while PRL uses progressive contrastive learning.
Both are evaluated with SASRec~(S) and BERT4Rec~(B) backbones.
\textbf{LARES}~\cite{lares2025} is a depth-recurrent reasoning model that iterates a shared core-block over the full token sequence, trained with self-supervised pre-training and reinforcement post-training.
ERL and PRL results come from the official ReaRec codebase; LARES from its official implementation.

\subsubsection{Implementation Details.}
We report NDCG@$K$ and Recall@$K$ for cutoffs 10 and 20, evaluated under full ranking.
To ensure fair comparison, all baselines use an embedding dimension of 256, a batch size of 2048, the AdamW optimizer at a learning rate of $10^{-3}$, early stopping with patience 10 on validation NDCG@10, and a temperature $\tau = 0.07$ across all experiments.

For RecRec Stage~1, the Context Compressor uses $R = 8$ latent interests, 1 layer, and 2 attention heads, with an FFN inner dimension of 512. $\lambda_{\text{IDR}}$ is tuned from $\{0, 0.01, 0.05, 0.1\}$.
For Stage~2, the reasoning configuration is $T = 3$ reasoning steps per supervision round, $K = 3$ supervision rounds, and $n = 3$ inner reasoning passes, giving $N = K \times T = 9$ reasoning steps at training.
The Reasoner uses 1 layer, 2 attention heads, and an FFN inner dimension of 512.

\subsection{Overall Performance (RQ1)}

\begin{table*}[!t]
\renewcommand{\arraystretch}{0.85}
\caption{Overall performance comparison.
  Best results are in \textbf{bold}, second best are \underline{underlined}. ``N@K'' and ``R@K'' denote ``NDCG@K'' and ``Recall@K'', respectively. $\ast$ denotes statistically significant improvement over the best baseline (paired $t$-test, $p < 0.05$). $\dagger$ marks reasoning-enhanced methods.
  S = SASRec backbone, B = BERT4Rec backbone.}
\label{tab:main_results}
\resizebox{\textwidth}{!}{%
\begin{tabular}{cl ccccccc >{\columncolor{gray!12}}c >{\columncolor{gray!12}}c}
\toprule
& & \multicolumn{2}{c}{Non-reasoning} & \multicolumn{5}{c}{Reasoning-enhanced$^\dagger$} & \multicolumn{2}{c}{Ours} \\
\cmidrule(lr){3-4} \cmidrule(lr){5-9} \cmidrule(lr){10-11}
Dataset & Metric & BERT4Rec & SASRec & ERL (B) & ERL (S) & PRL (B) & PRL (S) & LARES & RecRec (B) & RecRec (S) \\
\midrule
\multirow{4}{*}{S-Shop}
& N@10  & 0.0734 & 0.0737 & 0.0766 & 0.0776 & 0.0777 & 0.0818 & 0.0809 & \underline{0.0833} & \textbf{0.0853}$^\ast$ \\
& N@20  & 0.0950 & 0.0952 & 0.0982 & 0.0989 & 0.0997 & 0.1015 & 0.1009 & \underline{0.1039} & \textbf{0.1058}$^\ast$ \\
& R@10  & 0.1570 & 0.1544 & 0.1594 & 0.1616 & \underline{0.1627} & 0.1568 & 0.1530 & \textbf{0.1638} & 0.1619 \\
& R@20  & 0.2426 & 0.2399 & 0.2450 & \underline{0.2461} & \underline{0.2461} & 0.2351 & 0.2328 & \textbf{0.2501} & 0.2432 \\
\midrule
\multirow{4}{*}{V-Shop}
& N@10  & 0.0153 & 0.0158 & 0.0160 & 0.0161 & 0.0155 & 0.0173 & 0.0182 & \underline{0.0184} & \textbf{0.0197}$^\ast$ \\
& N@20  & 0.0196 & 0.0206 & 0.0204 & 0.0217 & 0.0215 & 0.0221 & 0.0222 & \underline{0.0241} & \textbf{0.0248}$^\ast$ \\
& R@10  & 0.0303 & 0.0289 & 0.0324 & 0.0378 & 0.0348 & 0.0381 & 0.0327 & \underline{0.0422} & \textbf{0.0449}$^\ast$ \\
& R@20  & 0.0476 & 0.0479 & 0.0500 & 0.0601 & 0.0586 & 0.0571 & 0.0488 & \underline{0.0649} & \textbf{0.0651}$^\ast$ \\
\midrule
\multirow{4}{*}{MicroLens}
& N@10  & 0.0428 & 0.0428 & 0.0435 & 0.0432 & 0.0448 & 0.0453 & 0.0460 & \underline{0.0481} & \textbf{0.0503}$^\ast$ \\
& N@20  & 0.0523 & 0.0533 & 0.0533 & 0.0538 & 0.0554 & 0.0562 & 0.0564 & \underline{0.0585} & \textbf{0.0611}$^\ast$ \\
& R@10  & 0.0885 & 0.0875 & 0.0865 & 0.0893 & 0.0898 & \underline{0.0908} & 0.0863 & 0.0921 & \textbf{0.0939} \\
& R@20  & 0.1263 & 0.1291 & 0.1251 & 0.1313 & 0.1318 & \underline{0.1344} & 0.1276 & 0.1331 & \textbf{0.1370} \\
\midrule
\multirow{4}{*}{Steam}
& N@10  & 0.0343 & 0.0347 & 0.0371 & 0.0383 & 0.0385 & 0.0408 & 0.0419 & \underline{0.0477} & \textbf{0.0481}$^\ast$ \\
& N@20  & 0.0459 & 0.0454 & 0.0496 & 0.0505 & 0.0501 & 0.0535 & 0.0545 & \underline{0.0607} & \textbf{0.0621}$^\ast$ \\
& R@10  & 0.0695 & 0.0698 & 0.0723 & 0.0780 & 0.0754 & 0.0804 & 0.0831 & \underline{0.0922} & \textbf{0.0978}$^\ast$ \\
& R@20  & 0.1158 & 0.1126 & 0.1219 & 0.1266 & 0.1214 & 0.1308 & 0.1330 & \underline{0.1440} & \textbf{0.1536}$^\ast$ \\
\bottomrule
\end{tabular}%
}
\end{table*}

Table~\ref{tab:main_results} presents the overall results.
For a fair comparison with baselines whose reasoning depth is fixed at training, we report RecRec's prediction at $N = 9$, without selecting the inference-time peak; Section~\ref{sec:rq2} examines those sweeps.
RecRec variants achieve the best performance on all four datasets.
RecRec~(S) improves NDCG@10 over the best reasoning-enhanced baseline by 14.8\% on Steam, 9.3\% on MicroLens, 8.2\% on V-Shop, and 4.3\% on S-Shop.
These gains come against baselines that either entangle the reasoning state with the input sequence or derive predictions from a single $d$-dim vector, both limitations that RecRec removes by design.
RecRec~(S) generally outperforms RecRec~(B), opposite to what one might expect from BERT4Rec's bidirectional context.
NDCG gains match or exceed Recall gains across datasets, indicating that recursive reasoning helps rank the target item higher rather than merely expanding the recall set.

\noindent\textbf{Answer to RQ1:}
RecRec's dual-state design outperforms all reasoning-enhanced baselines across four datasets.
These baselines either entangle reasoning with the input sequence or predict from a single vector; decoupling reasoning from prediction and using multiple latent interests both contribute to this lift.

\subsection{Inference-Time Reasoning Depth (RQ2)}
\label{sec:rq2}

\begin{figure*}[t]
  \centering
  \includegraphics[width=\linewidth]{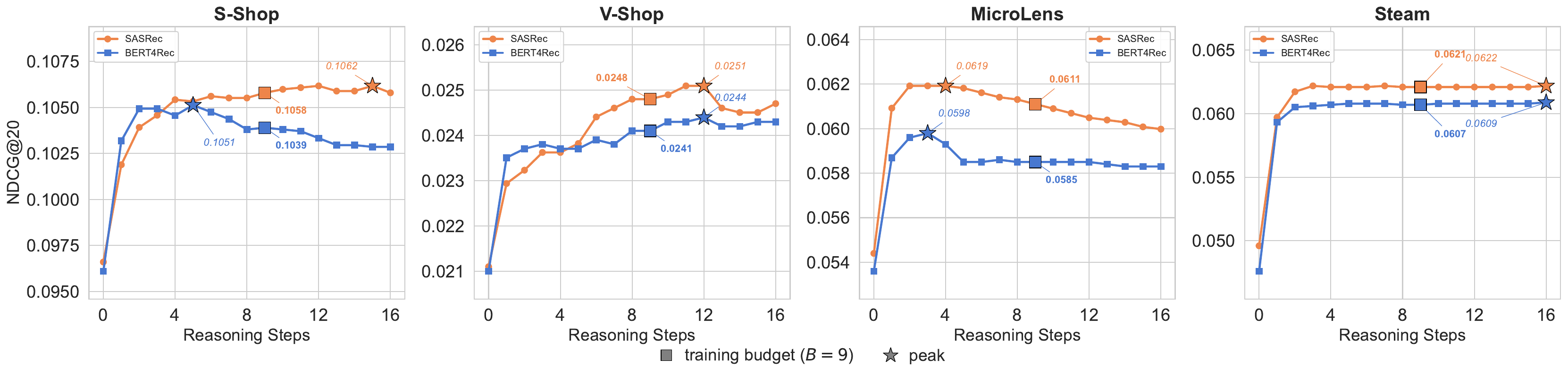}
  \caption{NDCG@20 as the inference-time $N$ is swept from $0$ to $16$ on each dataset.
    Squares mark $N=K\times T=9$ used at training and stars mark the best $N$ in the sweep.
    Both backbones start from the Stage-1 baseline at $N=0$ (no reasoning) and are free to use more reasoning steps at inference than they were trained with.}
  \label{fig:reasoning_depth}
\end{figure*}

Figure~\ref{fig:reasoning_depth} reports NDCG@20 as the inference-time $N$ varies from $0$ to $16$.
Two patterns stand out across datasets.
Moving from $N = 0$ (the Stage-1 baseline) to $N = 1$ produces a large jump in NDCG, for example from $0.0966$ to $0.103$ on S-Shop (SASRec) and from $0.0496$ to $0.058$ on Steam (SASRec).
The first Reasoner pass alone therefore accounts for a large share of the gain over the Stage-1 baseline.
Additional reasoning steps continue to refine the state, with diminishing returns.
The Reasoner improves a rough starting state substantially within a few passes, while refinement of an already-improved state produces smaller gains.

Increasing $N$ beyond $9$ continues to yield small additional improvements on S-Shop, V-Shop, and Steam, where the best-performing $N$ lies at $15$, $12$, and $16$ respectively (e.g., from $0.1058$ to $0.1062$ on S-Shop (SASRec); from $0.0248$ to $0.0251$ on V-Shop (SASRec)).
The model remains stable when $N$ is pushed well beyond $9$.
On MicroLens, the initial Reasoner pass also produces a clear improvement, but the gains are smaller, and the curve saturates near $N \approx 4$ and then declines.

\noindent\textbf{Answer to RQ2:}
Deep supervision enables RecRec to use flexible reasoning depth at inference.
The largest gain comes from the first few steps, and on three of four datasets, performance is sustained past the training-time depth $N = 9$ with small additional gains.
This extrapolation is what deep supervision with detached steps sets up.
Each round is trained to improve whatever state it receives, so the Reasoner can keep doing this when more steps are appended.

\subsection{Ablation Study (RQ3)}
\label{sec:ablation}

\begin{figure}[t]
  \centering
  \includegraphics[width=\linewidth]{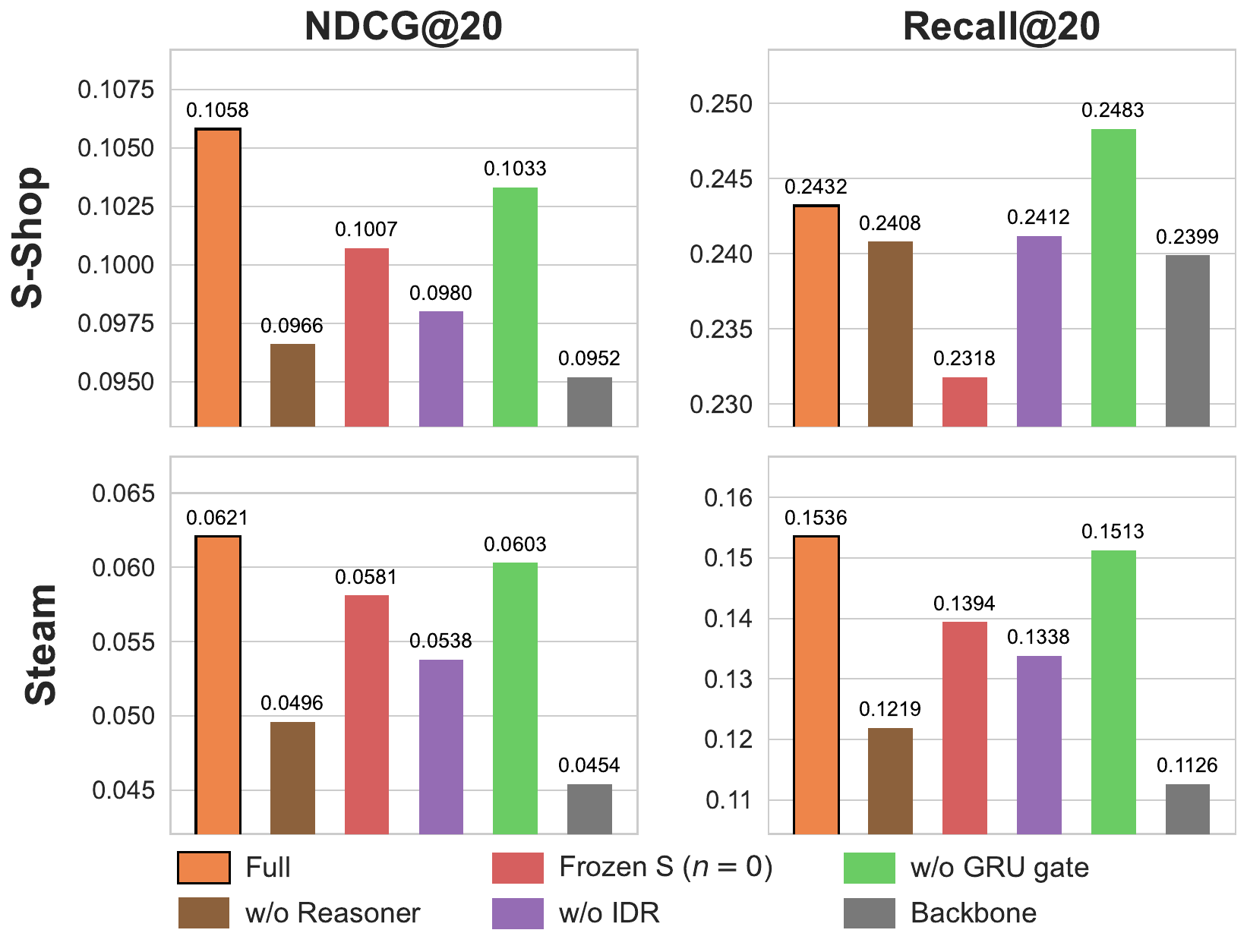}
  \caption{Ablation study of RecRec components on NDCG@20 and Recall@20 for S-Shop and Steam (SASRec backbone).
    \textbf{Full} is the complete RecRec; \textbf{w/o Reasoner} skips Stage-2; \textbf{Frozen S} ($n{=}0$) fixes the intermediate reasoning state; \textbf{w/o GRU gate} replaces the gated $\mathbf{M}$ update with overwriting; \textbf{w/o IDR} disables the diversity regularizer.}
  \label{fig:ablation_bars}
  \vspace{-0.5em}
\end{figure}

Figure~\ref{fig:ablation_bars} reports the ablation study on RecRec components on S-Shop and Steam with the SASRec backbone.
We evaluate four variants against the full RecRec and the raw SASRec backbone as a non-reasoning reference.
\emph{w/o Reasoner} predicts directly from $\mathbf{M}_0$, skipping Stage-2 entirely and testing the Reasoner's overall contribution.
\emph{Frozen $\mathbf{S}$ ($n{=}0$)} keeps $\mathbf{S}$ fixed at $\mathbf{S}_0$, isolating the iterative refinement of the intermediate reasoning state.
\emph{w/o GRU gate} replaces the gated $\mathbf{M}$ update with direct overwriting.
\emph{w/o IDR} removes the diversity regularizer from both training stages.
All variants are evaluated at $N = 9$.

\begin{figure}[!t]
  \centering
  \includegraphics[width=\linewidth]{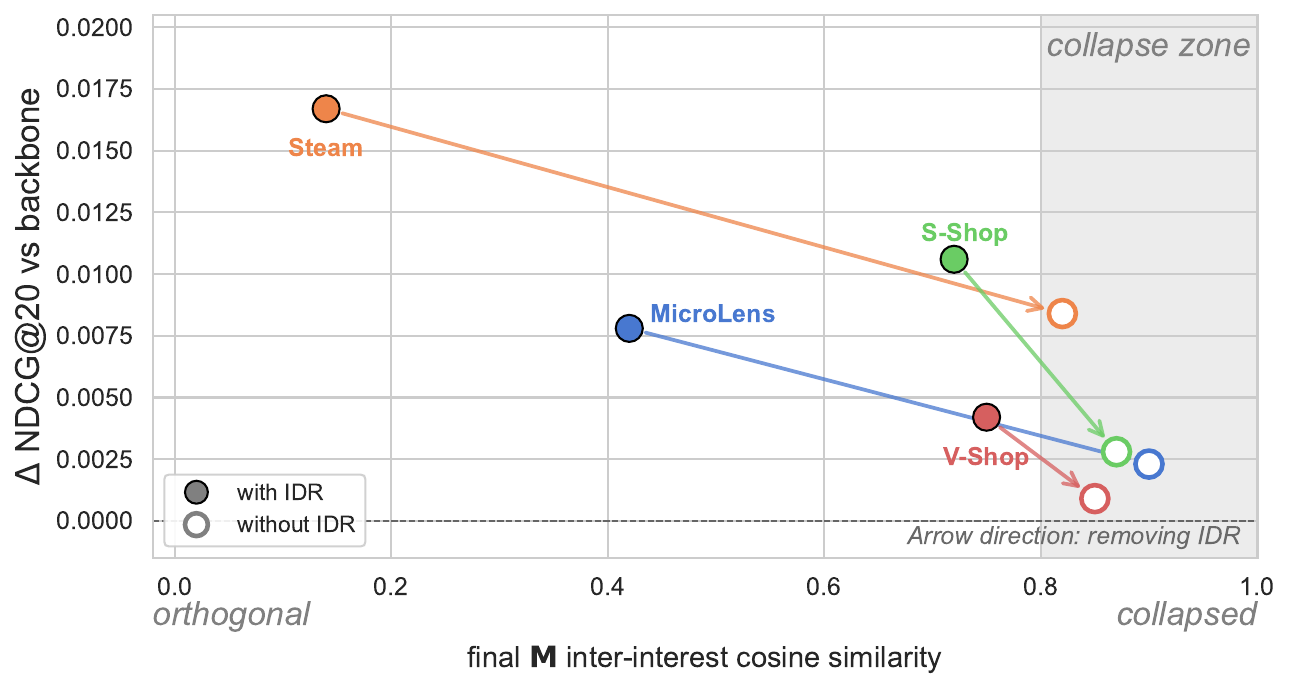}
  \caption{IDR mechanism: inter-interest cosine similarity of the final $\mathbf{M}$ versus NDCG@20 improvement over the SASRec backbone, on each dataset's test set.
    Filled circles are with IDR; hollow circles are without IDR; arrows show the effect of removing IDR.
    Without IDR, datasets enter the collapse zone (cosine $> 0.8$) and lose most of the IDR-induced gain.}
  \label{fig:idr_mechanism}
  \vspace{-1em}
\end{figure}

On both datasets, NDCG@20 follows the same pattern, with the raw SASRec backbone trailing all variants.
The full RecRec leads, and removing the Reasoner or IDR drops to roughly the Stage-1 baseline level while removing the GRU gate or fixing $\mathbf{S}$ at $\mathbf{S}_0$ causes moderate drops in between.
Removing the Reasoner equals the Stage-1 baseline by construction.
Removing IDR drops to a similar level: with the diversity regularizer disabled in both training stages, the $R$ latent interests collapse to near-identical directions in the final $\mathbf{M}$ (verified in Figure~\ref{fig:idr_mechanism}).
Fixing $\mathbf{S}$ at $\mathbf{S}_0$ (zero inner reasoning passes) lowers NDCG on both datasets, degenerating the dual-state design into $\mathbf{M}$-only evolution where the intermediate reasoning state contributes nothing $\mathbf{M}$ does not already carry.
Removing the GRU gate also costs NDCG, but less than fixing $\mathbf{S}$; the gate shapes how each candidate update is absorbed into $\mathbf{M}$ rather than whether the update mechanism works.
The Recall@20 ordering largely follows NDCG@20 with one exception.
On S-Shop, removing the gated update slightly exceeds full RecRec ($0.2483$ vs.\ $0.2432$), suggesting that the gate trades a small amount of Recall coverage for sharper ranking.

Figure~\ref{fig:idr_mechanism} measures the inter-interest cosine similarity of the final latent interests $\mathbf{M}$ on the test set.
Without IDR, $\mathbf{M}$'s $R$ interests collapse to nearly identical directions across all four datasets, falling into the marked collapse zone.
With IDR, the interests stay diverse to varying degrees across datasets.
The NDCG@20 improvement gained by enabling IDR (the vertical drop of each arrow) tracks the cosine reduction.
IDR keeps the $R$ interests separable, and the Reasoner's gain depends on having distinct interests to refine.

\paragraph{R sweep.}
Table~\ref{tab:r_sweep} reports the $R$ sweep on NDCG@20, with $R$ kept consistent between Stage~1 and Stage~2.
A single-vector latent ($R{=}1$) trails every multi-vector configuration on every dataset, the clearest signal that the prediction state benefits from carrying more than a single $d$-dimensional vector.
Beyond $R{=}2$ the optimum is dataset-dependent without a single explanatory axis.
$R{=}4$ wins on V-Shop and MicroLens, $R{=}8$ on Steam, $R{=}16$ on S-Shop.
Pushing to $R{=}16$ degrades performance on V-Shop and MicroLens, suggesting that an overly large $R$ can dilute the training signal across slots that the dataset cannot fully populate.
The default $R{=}8$ stays within $5\%$ of the per-dataset best, a safe middle-ground choice.

\begin{table}[!t]
\small
\caption{$R$ sweep: NDCG@20 with the SASRec backbone, $R$ kept consistent between Stage~1 and Stage~2.
Best is in \textbf{bold}.}
\label{tab:r_sweep}
\begin{tabular}{l ccccc}
\toprule
$R$ & 1 & 2 & 4 & 8 & 16 \\
\midrule
S-Shop    & 0.0990 & 0.1001 & 0.1078 & 0.1058 & \textbf{0.1095} \\
V-Shop & 0.0219 & 0.0237 & \textbf{0.0255} & 0.0248 & 0.0239 \\
MicroLens   & 0.0554 & 0.0614 & \textbf{0.0618} & 0.0611 & 0.0587 \\
Steam       & 0.0520 & 0.0567 & 0.0604 & \textbf{0.0621} & 0.0618 \\
\bottomrule
\end{tabular}
\vspace{-1em}
\end{table}

\noindent\textbf{Answer to RQ3:}
Every component of RecRec contributes.
The Reasoner, IDR, and $n$ inner reasoning passes that evolve $\mathbf{S}$ are required for the dual-state design to deliver, while the GRU gate shapes how each update is absorbed into $\mathbf{M}$ rather than whether the update works.
A multi-vector prediction latent ($R \geq 2$) outperforms a single vector across four datasets, with the optimal $R$ dataset-dependent.

\section{Conclusion}
\label{sec:conclusion}

We presented RecRec, a recursive reasoning framework for sequential recommendation that separates the reasoning state from the prediction state and predicts from $R$ latent interests rather than a single vector.
A Context Compressor with an Interest Diversity Regularizer initializes the latent interests, and a Recursive Reasoner refines them through alternating updates with a dedicated reasoning state.
The framework is trained in two supervised stages, with deep supervision and detached steps allowing inference-time reasoning depth to be adjusted freely.
Experiments on four datasets show consistent gains over reasoning-enhanced baselines; on three of four datasets, gains extend past the training-time reasoning depth.
Taken together, our results show that the gain from recursive reasoning in sequential recommendation comes primarily from two architectural choices: detaching the reasoning state from the prediction latent, and predicting from a set of latent interests rather than a single vector.
Both are lightweight and backbone-agnostic, suggesting that reasoning-state architecture is an orthogonal axis worth exploring alongside backbone improvements.

\paragraph{Future directions.}
A learned halting mechanism~\cite{jolicoeur2025less} would let RecRec stop reasoning early on easy sequences and continue on harder ones, but designing a halting signal for ranking objectives is non-trivial since there is no single discrete correct answer to supervise against.
Allowing asymmetric shapes for the reasoning and prediction states, currently constrained to the same $R{\times}d$ form, is a natural extension that would let the reasoning state carry more capacity than the prediction state.
A more behavioral study of how the latent interests evolve across reasoning steps, including gate dynamics and attention patterns, would clarify the mechanism behind the empirical lift and its variation across datasets.

\balance
\bibliographystyle{ACM-Reference-Format}
\bibliography{references}


\begin{thebibliography}{54}


\ifx \showCODEN    \undefined \def \showCODEN     #1{\unskip}     \fi
\ifx \showISBNx    \undefined \def \showISBNx     #1{\unskip}     \fi
\ifx \showISBNxiii \undefined \def \showISBNxiii  #1{\unskip}     \fi
\ifx \showISSN     \undefined \def \showISSN      #1{\unskip}     \fi
\ifx \showLCCN     \undefined \def \showLCCN      #1{\unskip}     \fi
\ifx \shownote     \undefined \def \shownote      #1{#1}          \fi
\ifx \showarticletitle \undefined \def \showarticletitle #1{#1}   \fi
\ifx \showURL      \undefined \def \showURL       {\relax}        \fi
\providecommand\bibfield[2]{#2}
\providecommand\bibinfo[2]{#2}
\providecommand\natexlab[1]{#1}
\providecommand\showeprint[2][]{arXiv:#2}

\bibitem[Bao et~al\mbox{.}(2023)]%
        {bao2023tallrec}
\bibfield{author}{\bibinfo{person}{Keqin Bao}, \bibinfo{person}{Jizhi Zhang},
  \bibinfo{person}{Yang Zhang}, \bibinfo{person}{Wenjie Wang},
  \bibinfo{person}{Fuli Feng}, {and} \bibinfo{person}{Xiangnan He}.}
  \bibinfo{year}{2023}\natexlab{}.
\newblock \showarticletitle{Tallrec: An effective and efficient tuning
  framework to align large language model with recommendation}. In
  \bibinfo{booktitle}{\emph{Proceedings of the 17th ACM conference on
  recommender systems}}. \bibinfo{pages}{1007--1014}.
\newblock


\bibitem[Bulatov et~al\mbox{.}(2022)]%
        {bulatov2022recurrent}
\bibfield{author}{\bibinfo{person}{Aydar Bulatov}, \bibinfo{person}{Yury
  Kuratov}, {and} \bibinfo{person}{Mikhail Burtsev}.}
  \bibinfo{year}{2022}\natexlab{}.
\newblock \showarticletitle{Recurrent memory transformer}.
\newblock \bibinfo{journal}{\emph{Advances in Neural Information Processing
  Systems}}  \bibinfo{volume}{35} (\bibinfo{year}{2022}),
  \bibinfo{pages}{11079--11091}.
\newblock


\bibitem[Carion et~al\mbox{.}(2020)]%
        {carion2020detr}
\bibfield{author}{\bibinfo{person}{Nicolas Carion}, \bibinfo{person}{Francisco
  Massa}, \bibinfo{person}{Gabriel Synnaeve}, \bibinfo{person}{Nicolas
  Usunier}, \bibinfo{person}{Alexander Kirillov}, {and} \bibinfo{person}{Sergey
  Zagoruyko}.} \bibinfo{year}{2020}\natexlab{}.
\newblock \showarticletitle{End-to-end object detection with transformers}. In
  \bibinfo{booktitle}{\emph{European conference on computer vision}}. Springer,
  \bibinfo{pages}{213--229}.
\newblock


\bibitem[Cen et~al\mbox{.}(2020)]%
        {cen2020comirec}
\bibfield{author}{\bibinfo{person}{Yukuo Cen}, \bibinfo{person}{Jianwei Zhang},
  \bibinfo{person}{Xu Zou}, \bibinfo{person}{Chang Zhou},
  \bibinfo{person}{Hongxia Yang}, {and} \bibinfo{person}{Jie Tang}.}
  \bibinfo{year}{2020}\natexlab{}.
\newblock \showarticletitle{Controllable multi-interest framework for
  recommendation}. In \bibinfo{booktitle}{\emph{Proceedings of the 26th ACM
  SIGKDD international conference on knowledge discovery \& data mining}}.
  \bibinfo{pages}{2942--2951}.
\newblock


\bibitem[Cho et~al\mbox{.}(2014)]%
        {cho2014learning}
\bibfield{author}{\bibinfo{person}{Kyunghyun Cho}, \bibinfo{person}{Bart
  Van~Merri{\"e}nboer}, \bibinfo{person}{{\c{C}}a{\u{g}}lar Gul{\c{c}}ehre},
  \bibinfo{person}{Dzmitry Bahdanau}, \bibinfo{person}{Fethi Bougares},
  \bibinfo{person}{Holger Schwenk}, {and} \bibinfo{person}{Yoshua Bengio}.}
  \bibinfo{year}{2014}\natexlab{}.
\newblock \showarticletitle{Learning phrase representations using RNN
  encoder--decoder for statistical machine translation}. In
  \bibinfo{booktitle}{\emph{Proceedings of the 2014 conference on empirical
  methods in natural language processing (EMNLP)}}.
  \bibinfo{pages}{1724--1734}.
\newblock


\bibitem[Dao and Gu(2024)]%
        {dao2024transformers}
\bibfield{author}{\bibinfo{person}{Tri Dao} {and} \bibinfo{person}{Albert Gu}.}
  \bibinfo{year}{2024}\natexlab{}.
\newblock \showarticletitle{Transformers are ssms: Generalized models and
  efficient algorithms through structured state space duality}.
\newblock \bibinfo{journal}{\emph{arXiv preprint arXiv:2405.21060}}
  (\bibinfo{year}{2024}).
\newblock


\bibitem[Dehghani et~al\mbox{.}(2018)]%
        {dehghani2019universal}
\bibfield{author}{\bibinfo{person}{Mostafa Dehghani}, \bibinfo{person}{Stephan
  Gouws}, \bibinfo{person}{Oriol Vinyals}, \bibinfo{person}{Jakob Uszkoreit},
  {and} \bibinfo{person}{{\L}ukasz Kaiser}.} \bibinfo{year}{2018}\natexlab{}.
\newblock \showarticletitle{Universal transformers}.
\newblock \bibinfo{journal}{\emph{arXiv preprint arXiv:1807.03819}}
  (\bibinfo{year}{2018}).
\newblock


\bibitem[Deng et~al\mbox{.}(2025)]%
        {deng2025rapo}
\bibfield{author}{\bibinfo{person}{Wenhao Deng}, \bibinfo{person}{Long Wei},
  \bibinfo{person}{Chenglei Yu}, {and} \bibinfo{person}{Tailin Wu}.}
  \bibinfo{year}{2025}\natexlab{}.
\newblock \showarticletitle{Unlocking reasoning capabilities in llms via
  reinforcement learning exploration}.
\newblock \bibinfo{journal}{\emph{arXiv preprint arXiv:2510.03865}}
  (\bibinfo{year}{2025}).
\newblock


\bibitem[Fu et~al\mbox{.}(2026)]%
        {fu2026diger}
\bibfield{author}{\bibinfo{person}{Junchen Fu}, \bibinfo{person}{Xuri Ge},
  \bibinfo{person}{Alexandros Karatzoglou}, \bibinfo{person}{Ioannis Arapakis},
  \bibinfo{person}{Suzan Verberne}, \bibinfo{person}{Joemon~M Jose}, {and}
  \bibinfo{person}{Zhaochun Ren}.} \bibinfo{year}{2026}\natexlab{}.
\newblock \showarticletitle{Differentiable Semantic ID for Generative
  Recommendation}.
\newblock \bibinfo{journal}{\emph{arXiv preprint arXiv:2601.19711}}
  (\bibinfo{year}{2026}).
\newblock


\bibitem[Fu et~al\mbox{.}(2024a)]%
        {fu2024iisan}
\bibfield{author}{\bibinfo{person}{Junchen Fu}, \bibinfo{person}{Xuri Ge},
  \bibinfo{person}{Xin Xin}, \bibinfo{person}{Alexandros Karatzoglou},
  \bibinfo{person}{Ioannis Arapakis}, \bibinfo{person}{Jie Wang}, {and}
  \bibinfo{person}{Joemon~M Jose}.} \bibinfo{year}{2024}\natexlab{a}.
\newblock \showarticletitle{IISAN: Efficiently adapting multimodal
  representation for sequential recommendation with decoupled PEFT}. In
  \bibinfo{booktitle}{\emph{Proceedings of the 47th International ACM SIGIR
  Conference on Research and Development in Information Retrieval}}.
  \bibinfo{pages}{687--697}.
\newblock


\bibitem[Fu et~al\mbox{.}(2025a)]%
        {fu2025efficient}
\bibfield{author}{\bibinfo{person}{Junchen Fu}, \bibinfo{person}{Xuri Ge},
  \bibinfo{person}{Xin Xin}, \bibinfo{person}{Alexandros Karatzoglou},
  \bibinfo{person}{Ioannis Arapakis}, \bibinfo{person}{Kaiwen Zheng},
  \bibinfo{person}{Yongxin Ni}, {and} \bibinfo{person}{Joemon M~Jose Joemon}.}
  \bibinfo{year}{2025}\natexlab{a}.
\newblock \showarticletitle{Efficient and effective adaptation of multimodal
  foundation models in sequential recommendation}.
\newblock \bibinfo{journal}{\emph{IEEE Transactions on Knowledge and Data
  Engineering}} (\bibinfo{year}{2025}).
\newblock


\bibitem[Fu et~al\mbox{.}(2025b)]%
        {fu2025crossan}
\bibfield{author}{\bibinfo{person}{Junchen Fu}, \bibinfo{person}{Yongxin Ni},
  \bibinfo{person}{Joemon~M Jose}, \bibinfo{person}{Ioannis Arapakis},
  \bibinfo{person}{Kaiwen Zheng}, \bibinfo{person}{Youhua Li}, {and}
  \bibinfo{person}{Xuri Ge}.} \bibinfo{year}{2025}\natexlab{b}.
\newblock \showarticletitle{Crossan: Towards efficient and effective adaptation
  of multiple multimodal foundation models for sequential recommendation}.
\newblock \bibinfo{journal}{\emph{arXiv preprint arXiv:2504.10307}}
  (\bibinfo{year}{2025}).
\newblock


\bibitem[Fu et~al\mbox{.}(2024b)]%
        {fu2024exploring}
\bibfield{author}{\bibinfo{person}{Junchen Fu}, \bibinfo{person}{Fajie Yuan},
  \bibinfo{person}{Yu Song}, \bibinfo{person}{Zheng Yuan},
  \bibinfo{person}{Mingyue Cheng}, \bibinfo{person}{Shenghui Cheng},
  \bibinfo{person}{Jiaqi Zhang}, \bibinfo{person}{Jie Wang}, {and}
  \bibinfo{person}{Yunzhu Pan}.} \bibinfo{year}{2024}\natexlab{b}.
\newblock \showarticletitle{Exploring adapter-based transfer learning for
  recommender systems: Empirical studies and practical insights}. In
  \bibinfo{booktitle}{\emph{Proceedings of the 17th ACM international
  conference on web search and data mining}}. \bibinfo{pages}{208--217}.
\newblock


\bibitem[Geiping et~al\mbox{.}(2025)]%
        {geiping2025scaling}
\bibfield{author}{\bibinfo{person}{Jonas Geiping}, \bibinfo{person}{Sean
  McLeish}, \bibinfo{person}{Neel Jain}, \bibinfo{person}{John Kirchenbauer},
  \bibinfo{person}{Siddharth Singh}, \bibinfo{person}{Brian~R Bartoldson},
  \bibinfo{person}{Bhavya Kailkhura}, \bibinfo{person}{Abhinav Bhatele}, {and}
  \bibinfo{person}{Tom Goldstein}.} \bibinfo{year}{2025}\natexlab{}.
\newblock \showarticletitle{Scaling up test-time compute with latent reasoning:
  A recurrent depth approach}.
\newblock \bibinfo{journal}{\emph{arXiv preprint arXiv:2502.05171}}
  (\bibinfo{year}{2025}).
\newblock


\bibitem[Geng et~al\mbox{.}(2022)]%
        {geng2022p5}
\bibfield{author}{\bibinfo{person}{Shijie Geng}, \bibinfo{person}{Shuchang
  Liu}, \bibinfo{person}{Zuohui Fu}, \bibinfo{person}{Yingqiang Ge}, {and}
  \bibinfo{person}{Yongfeng Zhang}.} \bibinfo{year}{2022}\natexlab{}.
\newblock \showarticletitle{Recommendation as language processing (rlp): A
  unified pretrain, personalized prompt \& predict paradigm (p5)}. In
  \bibinfo{booktitle}{\emph{Proceedings of the 16th ACM conference on
  recommender systems}}. \bibinfo{pages}{299--315}.
\newblock


\bibitem[Gu et~al\mbox{.}(2021)]%
        {gu2022efficiently}
\bibfield{author}{\bibinfo{person}{Albert Gu}, \bibinfo{person}{Karan Goel},
  {and} \bibinfo{person}{Christopher R{\'e}}.} \bibinfo{year}{2021}\natexlab{}.
\newblock \showarticletitle{Efficiently modeling long sequences with structured
  state spaces}.
\newblock \bibinfo{journal}{\emph{arXiv preprint arXiv:2111.00396}}
  (\bibinfo{year}{2021}).
\newblock


\bibitem[Guo et~al\mbox{.}(2026)]%
        {guo2026promise}
\bibfield{author}{\bibinfo{person}{Chengcheng Guo}, \bibinfo{person}{Kuo Cai},
  \bibinfo{person}{Yu Zhou}, \bibinfo{person}{Qiang Luo},
  \bibinfo{person}{Ruiming Tang}, \bibinfo{person}{Han Li},
  \bibinfo{person}{Kun Gai}, {and} \bibinfo{person}{Guorui Zhou}.}
  \bibinfo{year}{2026}\natexlab{}.
\newblock \showarticletitle{PROMISE: Process Reward Models Unlock Test-Time
  Scaling Laws in Generative Recommendations}.
\newblock \bibinfo{journal}{\emph{arXiv preprint arXiv:2601.04674}}
  (\bibinfo{year}{2026}).
\newblock


\bibitem[Guo et~al\mbox{.}(2025)]%
        {guo2025deepseek}
\bibfield{author}{\bibinfo{person}{Daya Guo}, \bibinfo{person}{Dejian Yang},
  \bibinfo{person}{Haowei Zhang}, \bibinfo{person}{Junxiao Song},
  \bibinfo{person}{Peiyi Wang}, \bibinfo{person}{Qihao Zhu},
  \bibinfo{person}{Runxin Xu}, \bibinfo{person}{Ruoyu Zhang},
  \bibinfo{person}{Shirong Ma}, \bibinfo{person}{Xiao Bi}, {et~al\mbox{.}}}
  \bibinfo{year}{2025}\natexlab{}.
\newblock \showarticletitle{Deepseek-r1: Incentivizing reasoning capability in
  llms via reinforcement learning}.
\newblock \bibinfo{journal}{\emph{arXiv preprint arXiv:2501.12948}}
  (\bibinfo{year}{2025}).
\newblock


\bibitem[Hao et~al\mbox{.}(2024)]%
        {hao2024coconut}
\bibfield{author}{\bibinfo{person}{Shibo Hao}, \bibinfo{person}{Sainbayar
  Sukhbaatar}, \bibinfo{person}{DiJia Su}, \bibinfo{person}{Xian Li},
  \bibinfo{person}{Zhiting Hu}, \bibinfo{person}{Jason Weston}, {and}
  \bibinfo{person}{Yuandong Tian}.} \bibinfo{year}{2024}\natexlab{}.
\newblock \showarticletitle{Training large language models to reason in a
  continuous latent space}.
\newblock \bibinfo{journal}{\emph{arXiv preprint arXiv:2412.06769}}
  (\bibinfo{year}{2024}).
\newblock


\bibitem[He et~al\mbox{.}(2025)]%
        {he2025double}
\bibfield{author}{\bibinfo{person}{Yaoqin He}, \bibinfo{person}{Junchen Fu},
  \bibinfo{person}{Kaiwen Zheng}, \bibinfo{person}{Songpei Xu},
  \bibinfo{person}{Fuhai Chen}, \bibinfo{person}{Jie Li},
  \bibinfo{person}{Joemon~M Jose}, {and} \bibinfo{person}{Xuri Ge}.}
  \bibinfo{year}{2025}\natexlab{}.
\newblock \showarticletitle{Double-filter: Efficient fine-tuning of pre-trained
  vision-language models via patch\&layer filtering}. In
  \bibinfo{booktitle}{\emph{Forty-second International Conference on Machine
  Learning}}.
\newblock


\bibitem[Hidasi et~al\mbox{.}(2015)]%
        {hidasi2016session}
\bibfield{author}{\bibinfo{person}{Bal{\'a}zs Hidasi},
  \bibinfo{person}{Alexandros Karatzoglou}, \bibinfo{person}{Linas Baltrunas},
  {and} \bibinfo{person}{Domonkos Tikk}.} \bibinfo{year}{2015}\natexlab{}.
\newblock \showarticletitle{Session-based recommendations with recurrent neural
  networks}.
\newblock \bibinfo{journal}{\emph{arXiv preprint arXiv:1511.06939}}
  (\bibinfo{year}{2015}).
\newblock


\bibitem[Hou et~al\mbox{.}(2024)]%
        {hou2024bridging}
\bibfield{author}{\bibinfo{person}{Yupeng Hou}, \bibinfo{person}{Jiacheng Li},
  \bibinfo{person}{Zhankui He}, \bibinfo{person}{An Yan},
  \bibinfo{person}{Xiusi Chen}, {and} \bibinfo{person}{Julian McAuley}.}
  \bibinfo{year}{2024}\natexlab{}.
\newblock \showarticletitle{Bridging language and items for retrieval and
  recommendation}.
\newblock \bibinfo{journal}{\emph{arXiv preprint arXiv:2403.03952}}
  (\bibinfo{year}{2024}).
\newblock


\bibitem[Hou et~al\mbox{.}(2022)]%
        {hou2022towards}
\bibfield{author}{\bibinfo{person}{Yupeng Hou}, \bibinfo{person}{Shanlei Mu},
  \bibinfo{person}{Wayne~Xin Zhao}, \bibinfo{person}{Yaliang Li},
  \bibinfo{person}{Bolin Ding}, {and} \bibinfo{person}{Ji-Rong Wen}.}
  \bibinfo{year}{2022}\natexlab{}.
\newblock \showarticletitle{Towards universal sequence representation learning
  for recommender systems}. In \bibinfo{booktitle}{\emph{Proceedings of the
  28th ACM SIGKDD conference on knowledge discovery and data mining}}.
  \bibinfo{pages}{585--593}.
\newblock


\bibitem[Jaegle et~al\mbox{.}(2021)]%
        {jaegle2021perceiver}
\bibfield{author}{\bibinfo{person}{Andrew Jaegle}, \bibinfo{person}{Felix
  Gimeno}, \bibinfo{person}{Andy Brock}, \bibinfo{person}{Oriol Vinyals},
  \bibinfo{person}{Andrew Zisserman}, {and} \bibinfo{person}{Joao Carreira}.}
  \bibinfo{year}{2021}\natexlab{}.
\newblock \showarticletitle{Perceiver: General perception with iterative
  attention}. In \bibinfo{booktitle}{\emph{International conference on machine
  learning}}. PMLR, \bibinfo{pages}{4651--4664}.
\newblock


\bibitem[Jiang et~al\mbox{.}(2026)]%
        {jiang2026diffureason}
\bibfield{author}{\bibinfo{person}{Jie Jiang}, \bibinfo{person}{Yang Wu},
  \bibinfo{person}{Qian Li}, \bibinfo{person}{Yuling Xiong},
  \bibinfo{person}{Yihang Su}, \bibinfo{person}{Junbang Huo},
  \bibinfo{person}{Longfei Lu}, \bibinfo{person}{Jun Zhang}, {and}
  \bibinfo{person}{Huan Yu}.} \bibinfo{year}{2026}\natexlab{}.
\newblock \showarticletitle{DiffuReason: Bridging Latent Reasoning and
  Generative Refinement for Sequential Recommendation}.
\newblock \bibinfo{journal}{\emph{arXiv preprint arXiv:2602.09744}}
  (\bibinfo{year}{2026}).
\newblock


\bibitem[Jolicoeur-Martineau(2025)]%
        {jolicoeur2025less}
\bibfield{author}{\bibinfo{person}{Alexia Jolicoeur-Martineau}.}
  \bibinfo{year}{2025}\natexlab{}.
\newblock \showarticletitle{Less is more: Recursive reasoning with tiny
  networks}.
\newblock \bibinfo{journal}{\emph{arXiv preprint arXiv:2510.04871}}
  (\bibinfo{year}{2025}).
\newblock


\bibitem[Kang and McAuley(2018)]%
        {kang2018self}
\bibfield{author}{\bibinfo{person}{Wang-Cheng Kang} {and}
  \bibinfo{person}{Julian McAuley}.} \bibinfo{year}{2018}\natexlab{}.
\newblock \showarticletitle{Self-attentive sequential recommendation}. In
  \bibinfo{booktitle}{\emph{2018 IEEE international conference on data mining
  (ICDM)}}. IEEE, \bibinfo{pages}{197--206}.
\newblock


\bibitem[Li et~al\mbox{.}(2019)]%
        {li2019mind}
\bibfield{author}{\bibinfo{person}{Chao Li}, \bibinfo{person}{Zhiyuan Liu},
  \bibinfo{person}{Mengmeng Wu}, \bibinfo{person}{Yuchi Xu},
  \bibinfo{person}{Huan Zhao}, \bibinfo{person}{Pipei Huang},
  \bibinfo{person}{Guoliang Kang}, \bibinfo{person}{Qiwei Chen},
  \bibinfo{person}{Wei Li}, {and} \bibinfo{person}{Dik~Lun Lee}.}
  \bibinfo{year}{2019}\natexlab{}.
\newblock \showarticletitle{Multi-interest network with dynamic routing for
  recommendation at Tmall}. In \bibinfo{booktitle}{\emph{Proceedings of the
  28th ACM international conference on information and knowledge management}}.
  \bibinfo{pages}{2615--2623}.
\newblock


\bibitem[Li et~al\mbox{.}(2023b)]%
        {li2023recformer}
\bibfield{author}{\bibinfo{person}{Jiacheng Li}, \bibinfo{person}{Ming Wang},
  \bibinfo{person}{Jin Li}, \bibinfo{person}{Jinmiao Fu}, \bibinfo{person}{Xin
  Shen}, \bibinfo{person}{Jingbo Shang}, {and} \bibinfo{person}{Julian
  McAuley}.} \bibinfo{year}{2023}\natexlab{b}.
\newblock \showarticletitle{Text is all you need: Learning language
  representations for sequential recommendation}. In
  \bibinfo{booktitle}{\emph{Proceedings of the 29th ACM SIGKDD Conference on
  Knowledge Discovery and Data Mining}}. \bibinfo{pages}{1258--1267}.
\newblock


\bibitem[Li et~al\mbox{.}(2025)]%
        {li2025exploring}
\bibfield{author}{\bibinfo{person}{Ruyu Li}, \bibinfo{person}{Wenhao Deng},
  \bibinfo{person}{Yu Cheng}, \bibinfo{person}{Zheng Yuan},
  \bibinfo{person}{Jiaqi Zhang}, {and} \bibinfo{person}{Fajie Yuan}.}
  \bibinfo{year}{2025}\natexlab{}.
\newblock \showarticletitle{Exploring the upper limits of text-based
  collaborative filtering using large language models: Discoveries and
  insights}. In \bibinfo{booktitle}{\emph{Proceedings of the 34th ACM
  International Conference on Information and Knowledge Management}}.
  \bibinfo{pages}{1643--1653}.
\newblock


\bibitem[Li et~al\mbox{.}(2023a)]%
        {li2024e4srec}
\bibfield{author}{\bibinfo{person}{Xinhang Li}, \bibinfo{person}{Chong Chen},
  \bibinfo{person}{Xiangyu Zhao}, \bibinfo{person}{Yong Zhang}, {and}
  \bibinfo{person}{Chunxiao Xing}.} \bibinfo{year}{2023}\natexlab{a}.
\newblock \showarticletitle{E4srec: An elegant effective efficient extensible
  solution of large language models for sequential recommendation}.
\newblock \bibinfo{journal}{\emph{arXiv preprint arXiv:2312.02443}}
  (\bibinfo{year}{2023}).
\newblock


\bibitem[Liao et~al\mbox{.}(2024)]%
        {liao2024llara}
\bibfield{author}{\bibinfo{person}{Jiayi Liao}, \bibinfo{person}{Sihang Li},
  \bibinfo{person}{Zhengyi Yang}, \bibinfo{person}{Jiancan Wu},
  \bibinfo{person}{Yancheng Yuan}, \bibinfo{person}{Xiang Wang}, {and}
  \bibinfo{person}{Xiangnan He}.} \bibinfo{year}{2024}\natexlab{}.
\newblock \showarticletitle{Llara: Large language-recommendation assistant}. In
  \bibinfo{booktitle}{\emph{Proceedings of the 47th International ACM SIGIR
  Conference on Research and Development in Information Retrieval}}.
  \bibinfo{pages}{1785--1795}.
\newblock


\bibitem[Liu et~al\mbox{.}(2025)]%
        {lares2025}
\bibfield{author}{\bibinfo{person}{Enze Liu}, \bibinfo{person}{Bowen Zheng},
  \bibinfo{person}{Xiaolei Wang}, \bibinfo{person}{Wayne~Xin Zhao},
  \bibinfo{person}{Jinpeng Wang}, \bibinfo{person}{Sheng Chen}, {and}
  \bibinfo{person}{Ji-Rong Wen}.} \bibinfo{year}{2025}\natexlab{}.
\newblock \showarticletitle{Lares: Latent reasoning for sequential
  recommendation}.
\newblock \bibinfo{journal}{\emph{arXiv preprint arXiv:2505.16865}}
  (\bibinfo{year}{2025}).
\newblock


\bibitem[Locatello et~al\mbox{.}(2020)]%
        {locatello2020object}
\bibfield{author}{\bibinfo{person}{Francesco Locatello}, \bibinfo{person}{Dirk
  Weissenborn}, \bibinfo{person}{Thomas Unterthiner}, \bibinfo{person}{Aravindh
  Mahendran}, \bibinfo{person}{Georg Heigold}, \bibinfo{person}{Jakob
  Uszkoreit}, \bibinfo{person}{Alexey Dosovitskiy}, {and}
  \bibinfo{person}{Thomas Kipf}.} \bibinfo{year}{2020}\natexlab{}.
\newblock \showarticletitle{Object-centric learning with slot attention}.
\newblock \bibinfo{journal}{\emph{Advances in neural information processing
  systems}}  \bibinfo{volume}{33} (\bibinfo{year}{2020}),
  \bibinfo{pages}{11525--11538}.
\newblock


\bibitem[Ni et~al\mbox{.}(2025)]%
        {ni2023content}
\bibfield{author}{\bibinfo{person}{Yongxin Ni}, \bibinfo{person}{Yu Cheng},
  \bibinfo{person}{Xiangyan Liu}, \bibinfo{person}{Junchen Fu},
  \bibinfo{person}{Youhua Li}, \bibinfo{person}{Xiangnan He},
  \bibinfo{person}{Yongfeng Zhang}, {and} \bibinfo{person}{Fajie Yuan}.}
  \bibinfo{year}{2025}\natexlab{}.
\newblock \showarticletitle{A content-driven micro-video recommendation dataset
  at scale}. In \bibinfo{booktitle}{\emph{Proceedings of the 34th ACM
  International Conference on Information and Knowledge Management}}.
  \bibinfo{pages}{6486--6491}.
\newblock


\bibitem[Qin et~al\mbox{.}(2025)]%
        {qin2025more}
\bibfield{author}{\bibinfo{person}{Weicong Qin}, \bibinfo{person}{Yi Xu},
  \bibinfo{person}{Weijie Yu}, \bibinfo{person}{Chenglei Shen},
  \bibinfo{person}{Xiao Zhang}, \bibinfo{person}{Ming He},
  \bibinfo{person}{Jianping Fan}, {and} \bibinfo{person}{Jun Xu}.}
  \bibinfo{year}{2025}\natexlab{}.
\newblock \showarticletitle{More: A mixture of reflectors framework for large
  language model-based sequential recommendation}. In
  \bibinfo{booktitle}{\emph{Proceedings of the Nineteenth ACM Conference on
  Recommender Systems}}. \bibinfo{pages}{299--308}.
\newblock


\bibitem[Rajput et~al\mbox{.}(2023)]%
        {rajput2023tiger}
\bibfield{author}{\bibinfo{person}{Shashank Rajput}, \bibinfo{person}{Nikhil
  Mehta}, \bibinfo{person}{Anima Singh}, \bibinfo{person}{Raghunandan
  Hulikal~Keshavan}, \bibinfo{person}{Trung Vu}, \bibinfo{person}{Lukasz
  Heldt}, \bibinfo{person}{Lichan Hong}, \bibinfo{person}{Yi Tay},
  \bibinfo{person}{Vinh Tran}, \bibinfo{person}{Jonah Samost}, {et~al\mbox{.}}}
  \bibinfo{year}{2023}\natexlab{}.
\newblock \showarticletitle{Recommender systems with generative retrieval}.
\newblock \bibinfo{journal}{\emph{Advances in Neural Information Processing
  Systems}}  \bibinfo{volume}{36} (\bibinfo{year}{2023}),
  \bibinfo{pages}{10299--10315}.
\newblock


\bibitem[Shao et~al\mbox{.}(2024)]%
        {shao2024deepseekmath}
\bibfield{author}{\bibinfo{person}{Zhihong Shao}, \bibinfo{person}{Peiyi Wang},
  \bibinfo{person}{Qihao Zhu}, \bibinfo{person}{Runxin Xu},
  \bibinfo{person}{Junxiao Song}, \bibinfo{person}{Xiao Bi},
  \bibinfo{person}{Haowei Zhang}, \bibinfo{person}{Mingchuan Zhang},
  \bibinfo{person}{YK Li}, \bibinfo{person}{Yang Wu}, {et~al\mbox{.}}}
  \bibinfo{year}{2024}\natexlab{}.
\newblock \showarticletitle{Deepseekmath: Pushing the limits of mathematical
  reasoning in open language models}.
\newblock \bibinfo{journal}{\emph{arXiv preprint arXiv:2402.03300}}
  (\bibinfo{year}{2024}).
\newblock


\bibitem[Shazeer(2020)]%
        {shazeer2020glu}
\bibfield{author}{\bibinfo{person}{Noam Shazeer}.}
  \bibinfo{year}{2020}\natexlab{}.
\newblock \showarticletitle{Glu variants improve transformer}.
\newblock \bibinfo{journal}{\emph{arXiv preprint arXiv:2002.05202}}
  (\bibinfo{year}{2020}).
\newblock


\bibitem[Sun et~al\mbox{.}(2019)]%
        {sun2019bert4rec}
\bibfield{author}{\bibinfo{person}{Fei Sun}, \bibinfo{person}{Jun Liu},
  \bibinfo{person}{Jian Wu}, \bibinfo{person}{Changhua Pei},
  \bibinfo{person}{Xiao Lin}, \bibinfo{person}{Wenwu Ou}, {and}
  \bibinfo{person}{Peng Jiang}.} \bibinfo{year}{2019}\natexlab{}.
\newblock \showarticletitle{BERT4Rec: Sequential recommendation with
  bidirectional encoder representations from transformer}. In
  \bibinfo{booktitle}{\emph{Proceedings of the 28th ACM international
  conference on information and knowledge management}}.
  \bibinfo{pages}{1441--1450}.
\newblock


\bibitem[Tang et~al\mbox{.}(2026)]%
        {tang2026parallel}
\bibfield{author}{\bibinfo{person}{Jiakai Tang}, \bibinfo{person}{Xu Chen},
  \bibinfo{person}{Wen Chen}, \bibinfo{person}{Jian Wu},
  \bibinfo{person}{Yuning Jiang}, {and} \bibinfo{person}{Bo Zheng}.}
  \bibinfo{year}{2026}\natexlab{}.
\newblock \showarticletitle{Parallel Latent Reasoning for Sequential
  Recommendation}.
\newblock \bibinfo{journal}{\emph{arXiv preprint arXiv:2601.03153}}
  (\bibinfo{year}{2026}).
\newblock


\bibitem[Tang et~al\mbox{.}(2025)]%
        {rearec2025}
\bibfield{author}{\bibinfo{person}{Jiakai Tang}, \bibinfo{person}{Sunhao Dai},
  \bibinfo{person}{Teng Shi}, \bibinfo{person}{Jun Xu}, \bibinfo{person}{Xu
  Chen}, \bibinfo{person}{Wen Chen}, \bibinfo{person}{Jian Wu}, {and}
  \bibinfo{person}{Yuning Jiang}.} \bibinfo{year}{2025}\natexlab{}.
\newblock \showarticletitle{Think before recommend: Unleashing the latent
  reasoning power for sequential recommendation}.
\newblock \bibinfo{journal}{\emph{arXiv preprint arXiv:2503.22675}}
  (\bibinfo{year}{2025}).
\newblock


\bibitem[Tang and Wang(2018)]%
        {tang2018personalized}
\bibfield{author}{\bibinfo{person}{Jiaxi Tang} {and} \bibinfo{person}{Ke
  Wang}.} \bibinfo{year}{2018}\natexlab{}.
\newblock \showarticletitle{Personalized top-n sequential recommendation via
  convolutional sequence embedding}. In \bibinfo{booktitle}{\emph{Proceedings
  of the eleventh ACM international conference on web search and data mining}}.
  \bibinfo{pages}{565--573}.
\newblock


\bibitem[Vaswani et~al\mbox{.}(2017)]%
        {vaswani2017attention}
\bibfield{author}{\bibinfo{person}{Ashish Vaswani}, \bibinfo{person}{Noam
  Shazeer}, \bibinfo{person}{Niki Parmar}, \bibinfo{person}{Jakob Uszkoreit},
  \bibinfo{person}{Llion Jones}, \bibinfo{person}{Aidan~N Gomez},
  \bibinfo{person}{{\L}ukasz Kaiser}, {and} \bibinfo{person}{Illia
  Polosukhin}.} \bibinfo{year}{2017}\natexlab{}.
\newblock \showarticletitle{Attention is all you need}.
\newblock \bibinfo{journal}{\emph{Advances in neural information processing
  systems}}  \bibinfo{volume}{30} (\bibinfo{year}{2017}).
\newblock


\bibitem[Wang et~al\mbox{.}(2025)]%
        {wang2025hierarchical}
\bibfield{author}{\bibinfo{person}{Guan Wang}, \bibinfo{person}{Jin Li},
  \bibinfo{person}{Yuhao Sun}, \bibinfo{person}{Xing Chen},
  \bibinfo{person}{Changling Liu}, \bibinfo{person}{Yue Wu},
  \bibinfo{person}{Meng Lu}, \bibinfo{person}{Sen Song}, {and}
  \bibinfo{person}{Yasin~Abbasi Yadkori}.} \bibinfo{year}{2025}\natexlab{}.
\newblock \showarticletitle{Hierarchical reasoning model}.
\newblock \bibinfo{journal}{\emph{arXiv preprint arXiv:2506.21734}}
  (\bibinfo{year}{2025}).
\newblock


\bibitem[Wei et~al\mbox{.}(2022)]%
        {wei2022chain}
\bibfield{author}{\bibinfo{person}{Jason Wei}, \bibinfo{person}{Xuezhi Wang},
  \bibinfo{person}{Dale Schuurmans}, \bibinfo{person}{Maarten Bosma},
  \bibinfo{person}{Fei Xia}, \bibinfo{person}{Ed Chi}, \bibinfo{person}{Quoc~V
  Le}, \bibinfo{person}{Denny Zhou}, {et~al\mbox{.}}}
  \bibinfo{year}{2022}\natexlab{}.
\newblock \showarticletitle{Chain-of-thought prompting elicits reasoning in
  large language models}.
\newblock \bibinfo{journal}{\emph{Advances in neural information processing
  systems}}  \bibinfo{volume}{35} (\bibinfo{year}{2022}),
  \bibinfo{pages}{24824--24837}.
\newblock


\bibitem[Yang et~al\mbox{.}(2026)]%
        {yang2026mancar}
\bibfield{author}{\bibinfo{person}{Kun Yang}, \bibinfo{person}{Yuxuan Zhu},
  \bibinfo{person}{Yazhe Chen}, \bibinfo{person}{Siyao Zheng},
  \bibinfo{person}{Bangyang Hong}, \bibinfo{person}{Kangle Wu},
  \bibinfo{person}{Yabo Ni}, \bibinfo{person}{Anxiang Zeng},
  \bibinfo{person}{Cong Fu}, {and} \bibinfo{person}{Hui Li}.}
  \bibinfo{year}{2026}\natexlab{}.
\newblock \showarticletitle{ManCAR: Manifold-Constrained Latent Reasoning with
  Adaptive Test-Time Computation for Sequential Recommendation}.
\newblock \bibinfo{journal}{\emph{arXiv preprint arXiv:2602.20093}}
  (\bibinfo{year}{2026}).
\newblock


\bibitem[Ye et~al\mbox{.}(2026)]%
        {ye2026multimodal}
\bibfield{author}{\bibinfo{person}{Yu Ye}, \bibinfo{person}{Junchen Fu},
  \bibinfo{person}{Yu Song}, \bibinfo{person}{Kaiwen Zheng}, {and}
  \bibinfo{person}{Joemon~M Jose}.} \bibinfo{year}{2026}\natexlab{}.
\newblock \showarticletitle{Are multimodal embeddings truly beneficial for
  recommendation? A deep dive into whole vs. individual modalities}. In
  \bibinfo{booktitle}{\emph{European Conference on Information Retrieval}}.
  Springer, \bibinfo{pages}{66--81}.
\newblock


\bibitem[Yu et~al\mbox{.}(2025)]%
        {yu2025dapo}
\bibfield{author}{\bibinfo{person}{Qiying Yu}, \bibinfo{person}{Zheng Zhang},
  \bibinfo{person}{Ruofei Zhu}, \bibinfo{person}{Yufeng Yuan},
  \bibinfo{person}{Xiaochen Zuo}, \bibinfo{person}{Yu Yue},
  \bibinfo{person}{Weinan Dai}, \bibinfo{person}{Tiantian Fan},
  \bibinfo{person}{Gaohong Liu}, \bibinfo{person}{Lingjun Liu},
  {et~al\mbox{.}}} \bibinfo{year}{2025}\natexlab{}.
\newblock \showarticletitle{Dapo: An open-source llm reinforcement learning
  system at scale}.
\newblock \bibinfo{journal}{\emph{arXiv preprint arXiv:2503.14476}}
  (\bibinfo{year}{2025}).
\newblock


\bibitem[Yuan et~al\mbox{.}(2023)]%
        {yuan2023go}
\bibfield{author}{\bibinfo{person}{Zheng Yuan}, \bibinfo{person}{Fajie Yuan},
  \bibinfo{person}{Yu Song}, \bibinfo{person}{Youhua Li},
  \bibinfo{person}{Junchen Fu}, \bibinfo{person}{Fei Yang},
  \bibinfo{person}{Yunzhu Pan}, {and} \bibinfo{person}{Yongxin Ni}.}
  \bibinfo{year}{2023}\natexlab{}.
\newblock \showarticletitle{Where to go next for recommender systems? id-vs.
  modality-based recommender models revisited}. In
  \bibinfo{booktitle}{\emph{Proceedings of the 46th International ACM SIGIR
  Conference on Research and Development in Information Retrieval}}.
  \bibinfo{pages}{2639--2649}.
\newblock


\bibitem[Zhai et~al\mbox{.}(2024)]%
        {zhai2024hstu}
\bibfield{author}{\bibinfo{person}{Jiaqi Zhai}, \bibinfo{person}{Lucy Liao},
  \bibinfo{person}{Xing Liu}, \bibinfo{person}{Yueming Wang},
  \bibinfo{person}{Rui Li}, \bibinfo{person}{Xuan Cao}, \bibinfo{person}{Leon
  Gao}, \bibinfo{person}{Zhaojie Gong}, \bibinfo{person}{Fangda Gu},
  \bibinfo{person}{Michael He}, {et~al\mbox{.}}}
  \bibinfo{year}{2024}\natexlab{}.
\newblock \showarticletitle{Actions speak louder than words: Trillion-parameter
  sequential transducers for generative recommendations}.
\newblock \bibinfo{journal}{\emph{arXiv preprint arXiv:2402.17152}}
  (\bibinfo{year}{2024}).
\newblock


\bibitem[Zhang et~al\mbox{.}(2022)]%
        {zhang2022re4}
\bibfield{author}{\bibinfo{person}{Shengyu Zhang}, \bibinfo{person}{Lingxiao
  Yang}, \bibinfo{person}{Dong Yao}, \bibinfo{person}{Yujie Lu},
  \bibinfo{person}{Fuli Feng}, \bibinfo{person}{Zhou Zhao},
  \bibinfo{person}{Tat-Seng Chua}, {and} \bibinfo{person}{Fei Wu}.}
  \bibinfo{year}{2022}\natexlab{}.
\newblock \showarticletitle{Re4: Learning to re-contrast, re-attend,
  re-construct for multi-interest recommendation}. In
  \bibinfo{booktitle}{\emph{Proceedings of the ACM web conference 2022}}.
  \bibinfo{pages}{2216--2226}.
\newblock


\bibitem[Zheng et~al\mbox{.}(2025)]%
        {zheng2025gspo}
\bibfield{author}{\bibinfo{person}{Chujie Zheng}, \bibinfo{person}{Shixuan
  Liu}, \bibinfo{person}{Mingze Li}, \bibinfo{person}{Xiong-Hui Chen},
  \bibinfo{person}{Bowen Yu}, \bibinfo{person}{Chang Gao}, \bibinfo{person}{Kai
  Dang}, \bibinfo{person}{Yuqiong Liu}, \bibinfo{person}{Rui Men},
  \bibinfo{person}{An Yang}, {et~al\mbox{.}}} \bibinfo{year}{2025}\natexlab{}.
\newblock \showarticletitle{Group sequence policy optimization}.
\newblock \bibinfo{journal}{\emph{arXiv preprint arXiv:2507.18071}}
  (\bibinfo{year}{2025}).
\newblock


\bibitem[Zhuang et~al\mbox{.}(2025)]%
        {zhuang2025frequency}
\bibfield{author}{\bibinfo{person}{Ziyi Zhuang}, \bibinfo{person}{Hongji Li},
  \bibinfo{person}{Junchen Fu}, \bibinfo{person}{Jiacheng Liu},
  \bibinfo{person}{Joemon~M Jose}, \bibinfo{person}{Youhua Li}, {and}
  \bibinfo{person}{Yongxin Ni}.} \bibinfo{year}{2025}\natexlab{}.
\newblock \showarticletitle{Frequency-Decoupled distillation for efficient
  multimodal recommendation}. In \bibinfo{booktitle}{\emph{Proceedings of the
  34th ACM International Conference on Information and Knowledge Management}}.
  \bibinfo{pages}{4571--4581}.
\newblock


\end{thebibliography}

\end{document}